\documentclass[sigconf]{acmart}
\usepackage[skip=0pt]{caption}
\usepackage{algorithmicx}
\usepackage{algorithm}
\usepackage[noend]{algpseudocode}
\usepackage{subcaption}
\usepackage{multirow}
\usepackage{xspace}
\usepackage{tikz}

\newcommand{\DCName}{\texttt{Helios}\xspace}
\newcommand{\Imp}[2]{\vspace{3pt}\noindent\fbox{\parbox{0.97\linewidth}{\textbf{Implication \##1:} #2}}\vspace{3pt}}
\newcommand\encircle[1]{\tikz[baseline=(X.base)] 
    \node (X) [draw, shape=circle, inner sep=0pt, fill=black, text=white] {\strut #1};}

\algnewcommand\algorithmicinput{\textbf{Input:}}
\algnewcommand\Input{\item[\algorithmicinput]}
\algnewcommand\algorithmicoutput{\textbf{Output:}}
\algnewcommand\Output{\item[\algorithmicoutput]}

\AtBeginDocument{%
  \providecommand\BibTeX{{%
    \normalfont B\kern-0.5em{\scshape i\kern-0.25em b}\kern-0.8em\TeX}}}

\copyrightyear{2021} 
\acmYear{2021} 
\setcopyright{acmcopyright}\acmConference[SC '21]{The International Conference for High Performance Computing, Networking, Storage and Analysis}{November 14--19, 2021}{St. Louis, MO, USA}
\acmBooktitle{The International Conference for High Performance Computing, Networking, Storage and Analysis (SC '21), November 14--19, 2021, St. Louis, MO, USA}
\acmPrice{15.00}
\acmDOI{10.1145/3458817.3476223}
\acmISBN{978-1-4503-8442-1/21/11}

\acmSubmissionID{pap594}

\begin{document}

\title{Characterization and Prediction of Deep Learning Workloads in Large-Scale GPU Datacenters}


\author{Qinghao Hu$^{1,2}$ \quad Peng Sun$^3$ \quad Shengen Yan$^3$ \quad Yonggang Wen$^1$ \quad Tianwei Zhang$^1$}
\affiliation{%
  \institution{$^1$School of Computer Science and Engineering, Nanyang Technological University}
  \department{$^2$S-Lab, Nanyang Technological University \quad \quad $^3$SenseTime}
  \country{\{qinghao.hu, ygwen, tianwei.zhang\}@ntu.edu.sg \quad \{sunpeng1, yanshengen\}@sensetime.com}}
\authornote{Corresponding author.}






%
\renewcommand{\shortauthors}{Qinghao Hu, Peng Sun, Shengen Yan, Yonggang Wen, and Tianwei Zhang}

\renewcommand{\authors}{Qinghao Hu, Peng Sun, Shengen Yan, Yonggang Wen, and Tianwei Zhang}

\begin{abstract}
  Modern GPU datacenters are critical for delivering Deep Learning (DL) models and services in both the research community and industry. When operating a datacenter, optimization of resource scheduling and management can bring significant financial benefits. Achieving this goal requires a deep understanding of the job features and user behaviors. We present a comprehensive study about the characteristics of DL jobs and resource management.
  First, we perform a large-scale analysis of real-world job traces from SenseTime. We uncover some interesting conclusions from the perspectives of clusters, jobs and users, which can facilitate the cluster system designs.
  Second, we introduce a general-purpose framework, which manages resources based on historical data. As case studies, we design (1) a \textsl{Quasi-Shortest-Service-First} scheduling service, which can minimize the cluster-wide average job completion time by up to 6.5$\times$; (2) a \textsl{Cluster Energy Saving} service, which improves overall cluster utilization by up to 13\%.

\end{abstract}

%
%


%

\maketitle

\section{Introduction}
Over the years, we have witnessed the remarkable impact of Deep Learning (DL) technology and applications on every aspect of our daily life, e.g., face recognition \cite{DeepFace}, language translation \cite{Seq2Seq}, advertisement recommendation \cite{YouTubeRecommender}, etc.
The outstanding performance of DL models comes from the complex neural network structures and may contain trillions of parameters \cite{fedus2021switch}. Training a production model may require large amounts of GPU resources to support thousands of petaflops operations \cite{GPT-3}. Hence, it is a common practice for research institutes, AI companies and cloud providers to build large-scale GPU clusters to facilitate DL model development. These clusters are managed in a multi-tenancy fashion, offering services to different groups and users based on their demands, with resource regulation and access controls.

A job scheduler is necessary to manage resources and schedule jobs. It determines the resource utilization of the entire cluster, and job performance, which further affects the operation cost and user experience. Understanding the characteristics of DL workloads is indispensable for managing and operating GPU clusters. DL jobs share some similar features as conventional HPC workloads, which are generally different from big data jobs in cloud computing (e.g., MapReduce). (1) \textsl{Iterative process} \cite{Themis,Optimus}. Similar to some numerical simulation and analysis jobs in HPC, typical DL training jobs are also iterative computation. The target model is obtained through the gradient descent update iteratively and the long-term training process can be suspended and resumed via checkpoints. (2) \textsl{Gang scheduling} \cite{gang}. DL training jobs require all the GPUs to be allocated simultaneously in an all-or-nothing manner \cite{Gandivafair,AntMan}. This may cause resource fragmentation in GPU clusters. (3) \textsl{Exclusive allocation} \cite{Philly,Tiresias}. The GPU resources are allocated exclusively in DL datacenters. Although the advanced NVIDIA Multi-Instance GPU (MIG) technology \cite{mig} provides intrinsic hardware-level support for fine-grained sharing on NVIDIA A100 GPUs, existing GPU datacenters built with previous-generation GPUs typically only support coarse-grained GPU allocation \cite{Salus}.

Furthermore, DL training jobs also exhibit some unique features different from most HPC or big data workloads. (1) \textsl{Inherent heterogeneity} \cite{Gandiva}. DL training jobs typically require a number of GPUs as the dominant resources, as well as associated resources such as CPUs and memory. (2) \textsl{Placement sensitivity} \cite{Themis}. Multi-GPU jobs usually perform gradient synchronization to update model parameters at the end of each iteration. Better interconnect topology can achieve lower communication overhead and faster training speed for multi-GPU jobs. Meanwhile, colocating multiple jobs in one server could cause performance degradation since the interference of system resources like PCIe bandwidth \cite{Philly}. (3) \textsl{Feedback-driven exploration} \cite{Gandiva}. To train a model, users typically try several configurations, and use early feedback from these jobs to decide which one should be kept or killed. This gives higher cancellation rates in GPU clusters \cite{Philly}.


A variety of works proposed new schedulers specifically for GPU clusters to achieve higher performance, resource utilization and fairness \cite{Themis, Gavel, HiveD, Tiresias, AntMan}. To adapt to the characteristics of DL workloads, these works adopt the Philly trace \cite{Philly} from Microsoft GPU cluster for design and evaluation. To the best of our knowledge, the Philly trace is the only publicly available DL workload dataset so far. However, analysis of only this trace may be inadequate to prove the generality of the designs and can cause the overfitting issue, as pointed in \cite{diversity,TraceSC20}. In addition, this trace was collected in 2017. Due to the rapid development of DL technology and demands, this trace may not be able to reflect the latest characteristics of DL jobs anymore.

In this paper, we present an in-depth study about the characterization of DL workloads and scheduler designs. Our study is based on a new set of DL job traces collected from a datacenter in SenseTime, named \DCName. These traces have the following benefits. (1) They were collected over six months in 2020, which can represent the emerging DL algorithms and models. (2) They incorporate more than 3 million jobs among four independent GPU clusters. These jobs exhibit high variety, covering different types (training, inference, data preprocessing, etc.), applications (computer vision, natural language processing, etc.), and purposes (product development, research etc.). (3) These traces include rich information, which enables thorough analysis from different aspects, and diverse evaluations of scheduling systems. It is worth noting that we do not aim to compare our traces with Philly and show the advantages. Instead, we release our traces and expect they can be an alternative to Philly. Our traces together with Philly can increase the diversity of job workloads to enable more general design and evaluations (as we will do for our prediction framework in this paper). We also hope this can inspire other organizations to release their job traces to benefit the community of deep learning systems.

Based on the traces, this paper makes two key contributions. First, we perform a large-scale analysis of the DL jobs in the four clusters, from the perspectives of jobs, clusters and users.
Prior works \cite{Philly, AliPAI} only focused on the job-level characterization. Our study provides more extensive analysis about the behaviors of GPU clusters and users. 
Through our analysis, we identify seven new implications, which can shed light on the design of GPU clusters.


Second, we introduce a novel prediction-based framework to manage compute resources and schedule DL jobs. This framework is inspired by some implications from our trace analysis: the characteristics of clusters and jobs exhibit obvious predictable patterns, which gives us the opportunities to forecast those behaviors in advance and then optimize the management of jobs and resources. This framework can be integrated with different services. Each service builds a machine learning model from the historical data, and predicts the future states of the cluster, or upcoming job information. Based on the prediction results, the service can make optimized actions for resource and job management. To maintain prominent performance, the prediction model is also kept updated with new data. We present two services as case studies: (1) A \textsl{Quasi-Shortest-Service-First} scheduling service can assign each new job a priority score based on the historical information. It then schedules the jobs based on these priority scores, which can reduce the queuing delay by up to 20.2$\times$ and improve the overall job completion time (JCT) by up to 6.5$\times$. (2) A \textsl{Cluster Energy Saving} service can proactively predict the demanded compute nodes in advance, and then leverages the Dynamic Resource Sleep (DRS) technique to efficiently power off the unnecessary nodes. It can improve up to 13\% of node utilization rate and conserve millions of kilowatt hours of electricity annually across the four clusters.

\begin{table}[t]
  \caption{Configurations of four clusters in \DCName (All data are collected on September 1st, 2020, except \# of jobs and VCs, which cover the period of April-September, 2020).}
  \label{tab_trace}
  \resizebox{1\linewidth}{!}{
    \begin{tabular}{@{}lccccc@{}}
      \toprule
                  & \textbf{Venus}                             & \textbf{Earth}                             & \textbf{Saturn} & \textbf{Uranus} & \textbf{Total} \\ \midrule
      CPU         & \multicolumn{2}{c}{Intel, 48 threads/node} & \multicolumn{2}{c}{Intel, 64 threads/node} & -                                                  \\
      RAM         & \multicolumn{2}{c}{376GB per node}         & \multicolumn{2}{c}{256GB per node}         & -                                                  \\
      Network     & \multicolumn{2}{c}{IB EDR}                 & \multicolumn{2}{c}{IB FDR}                 & -                                                  \\
      GPU model   & Volta                                      & Volta                                      & Pascal \& Volta & Pascal          & -              \\ \midrule
      \# of VCs   & 27                                         & 25                                         & 28              & 25              & 105            \\
      \# of Nodes & 133                                        & 143                                        & 262             & 264             & 802            \\
      \# of GPUs  & 1,064                                      & 1,144                                      & 2,096           & 2,112           & 6,416          \\
      \# of Jobs  & 247k                                       & 873k                                       & 1,753k          & 490k            & 3,363k         \\ \bottomrule
    \end{tabular}}
\end{table}

\section{Background}
\label{sec_background}
In this section, we first introduce our GPU datacenter, dubbed \DCName (\S\ref{ssec_architecture}). Then we describe the DL workloads running in this datacenter (\S\ref{ssec_workload}) and the corresponding job traces (\S\ref{ssec_dataset}).

\subsection{\DCName Datacenter}
\label{ssec_architecture}
\DCName is a private datacenter dedicated to developing DL models for research and production in SenseTime, containing  multiple multi-tenant clusters. In this paper, we select 4 representative clusters: \texttt{Venus}, \texttt{Earth}, \texttt{Saturn}, and \texttt{Uranus}. Table \ref{tab_trace} shows the configurations of each cluster. Note that \texttt{Saturn} is a heterogeneous cluster with mixed NVIDIA Pascal and Volta GPUs, while the other three clusters are composed of identical Volta or Pascal GPUs.  Our conclusions from the analysis of these 4 clusters are general for other clusters in \DCName as well.



Each cluster in \DCName serves multiple groups in SenseTime concurrently. To support resource isolation and management for multi-tenancy, a cluster is further divided into several Virtual Clusters (VCs), and each VC is dedicated to one group with its demanded resources. All GPUs within one VC are homogeneous. Each node is exclusively allocated to one VC, and over-subscription of GPU resource quota is not enabled in \DCName. Configuration of a VC can be dynamically changed in three situations: (1) when the demand of a group is increased, new servers will be purchased and allocated to its VC, or form a new VC; (2) when a group is less busy, the size of its VC may be scaled down; (3) when groups are combined or split, their VCs are also merged or split correspondingly.

To reduce the communication overhead in distributed workloads, GPUs are interconnected to each other via a hierarchic network: (1) intra-node communication is achieved via the high-bandwidth and low-latency PCIe (for Pascal GPUs) or NVLink (for Volta GPUs) \cite{nvlink} ; (2) inter-node communication within the same RDMA domain is achieved via the high-speed InfiniBand. To improve the workload performance and reduce network interference, cross-RDMA-domain distributed training (communication through TCP/IP) are not allowed in \DCName.

The distributed storage system is also critical for workload performance. To support the massive data throughput in DL jobs, \DCName adopts Lustre \cite{lustre} as the file system, and the input data for the jobs are stored in Ceph \cite{Ceph} as the object storage. In addition, Memcached \cite{Memcached} is used to accelerate data access.

The \emph{Slurm} workload manager \cite{SLURM} is adopted to regulate the resources and job execution. Specifically, it dynamically adjusts the configurations of VCs, including the resource quota in each VC, total number of VCs, job time limit, etc. Meanwhile, it is also responsible for the online scheduling of DL jobs, following three steps. (1) A user submits his or her job to a VC with specified job resource demands (e.g., numbers of GPUs and CPUs). If the CPU requirement is not specified, the scheduler will allocate CPU cores proportional to the requested GPU counts. (2) \emph{Slurm} maintains a separate allocation queue for each VC (\emph{VCQueue}), and selects jobs for scheduling.
All jobs in one \emph{VCQueue} have the same priority setting so the scheduling order is only determined by the job's submission time. (3) \emph{Slurm} allocates jobs in a consolidated paradigm by packing jobs into as few nodes as possible. A user can also select specific nodes if he or she has special topology requirements. For instance, some exploratory jobs are placed across specific numbers of nodes for testing the performance impact of GPU affinity. Only 0.15\% of the jobs are placed in a way customized by the users. After the job is scheduled, it keeps running until completion or being terminated by the user. Preemption is not supported in \DCName.

\vspace{-3pt}
\subsection{Workloads in \DCName}
\label{ssec_workload}
\DCName supports various types of jobs in the DL development pipeline, e.g., data preprocessing, model training, inference, quantization, etc.
These workloads are submitted by product groups for developing commercial products, as well as research groups for exploring new technologies. They range over different DL domains, including computer vision, natural language processing, reinforcement learning, etc.

A majority of the GPU jobs are DL training, which mainly follow an iterative fashion \cite{Tiresias}: the training task consists of many iterations, where gradient descent is applied to update model parameters based on the mini-batch data in each iteration. To scale with complex models and large datasets, many jobs adopt the distributed data-parallel training scheme across multiple GPUs. In each iteration, every GPU processes a subset of data in parallel and then performs gradient synchronization to update model parameters. This synchronization typically adopts the parameter sever \cite{PS} or all-reduce strategy for high-speed multi-node/multi-GPU communication (e.g., NCCL \cite{nccl} as backend). Users mainly adopt the built-in libraries (e.g., DistributedDataParallel in Pytorch, MultiWorkerMirroredStrategy in Tensorflow) to implement their jobs.

In addition, there are also a quantity of jobs for data/model preprocessing and postprocessing.  For instance, some CPU jobs generate large-scale training datasets by extracting frames from videos; some jobs rescale the images according to the model's requirements. To speed up model inference, it is common to perform post-training quantization to reduce model size before deployment.


\subsection{DL Job Traces from \DCName}
\label{ssec_dataset}
We collect jobs from each of the 4 clusters in \DCName, which serves as the basis of our analysis and system design in this paper. These four traces span 6 months from April 2020 to September 2020, covering a total of 3.36 million jobs over 802 compute nodes with 6416 GPUs. Each trace contains two parts: (1) We collect the job logs through the \emph{Slurm} \texttt{sacct} command, which provides the rich information for each job. (2) The daily VC configurations of each cluster from  \emph{Slurm}. Besides, we leverage node allocation details from the job logs to infer the timing information for each cluster.

To the best of our knowledge, this is the largest set of DL job traces, and also the first one with comprehensive types of jobs in addition to DL training. We release these traces to the research community, and expect they can benefit researchers for DL workload analysis and design of GPU datacenter systems.


\begin{table}[t]
  \caption{Comparisons between \DCName and Philly traces.}
  \label{tab_philly}
  \resizebox{\linewidth}{!}{
    \begin{tabular}{@{}lcc||lcc@{}}
      \toprule
                     & \textbf{Helios} & \textbf{Philly} &                    & \textbf{Helios} & \textbf{Philly} \\ \midrule
      \# of clusters & 4               & 1               & Duration           & 6 months        & 83 days         \\
      \# of VCs      & 105             & 14              & Average \# of GPUs & 3.72            & 1.75            \\
      \# of Jobs     & 3.36M           & 103k            & Average Duration   & 6,652s          & 28,329s         \\
      \# of GPU Jobs & 1.58M           & 103k            & Maximum \# of GPUs & 2,048           & 128             \\
      \# of CPU Jobs & 1.78M           & 0               & Maximum Duration   & 50 days         & 60 days         \\ \bottomrule
    \end{tabular}}
\end{table}

\subsubsection{Terminology}
\label{ssec_terminology}
We emphasize some terminologies in the job traces, which will be widely mentioned in our following analysis.

\textbf{Job status}: a job can end up with one of five statuses: (1) completed: it is finished successfully; (2) canceled: it is terminated by the user; (3) failed: it is terminated due to internal or external errors; (4) timeout: the execution time is out of limit; (5) node fail: it is terminated due to the node crash. Timeout and node fail are very rare in our traces, and will be regarded as failed in this study.

\textbf{CPU job}: this job is executed without any GPUs (e.g., image preprocessing, file decompression).

\textbf{GPU job}: the job needs to be executed on GPUs for acceleration (e.g., DL model training, model evaluation).

\textbf{GPU time}: this metric is used to quantify the amount of GPU resources required by the job. It is calculated as the product of total execution time and the number of GPUs.

\textbf{CPU time:} this is the product of total execution time and the number of CPUs. It is only considered for CPU job analysis.

\textbf{Cluster utilization}: this metric is used to characterize the resource utilization of a cluster. Since GPUs are the dominant resources in DL jobs, we calculate the cluster utilization as the ratio of active GPUs among the total GPUs in the cluster.


\subsubsection{Comparisons with the Philly Trace}
\label{ssec_philly}
Microsoft released a trace of DL training jobs from its internal cluster Philly \cite{Philly}. It is currently the most popular public trace containing rich information about production-level DL training jobs. A quantity of works on GPU resource management and optimization leveraged this Philly trace for analysis and evaluation \cite{Themis, Gavel, HiveD, Tiresias, AntMan, Pollux, AFS}.

DL has experienced rapid development over the years. New models and algorithms are emerging with increased resource demands \cite{GPT-3, fedus2021switch}. There is a 10.5$\times$ year-by-year increase in the number of DL jobs in Microsoft datacenters \cite{Tiresias}. Hence, the Philly trace collected in 2017 may not be able to accurately reflect the characteristics of modern GPU clusters and DL jobs. We expect our trace can fill this gap with a larger number of jobs and the latest DL models and algorithms. We present detailed comparisons between our \DCName trace and Microsoft Philly trace. Although the Philly trace ranges from August 2017 to December 2017, jobs in the first two months exhibit abnormal behaviors (much less density and different features). So we select the period of October 2017 to December 2017 with 103,467 jobs, which was also adopted in the original paper \cite{Philly}.

\begin{figure}[t]
  \centering
  \includegraphics[width=\linewidth]{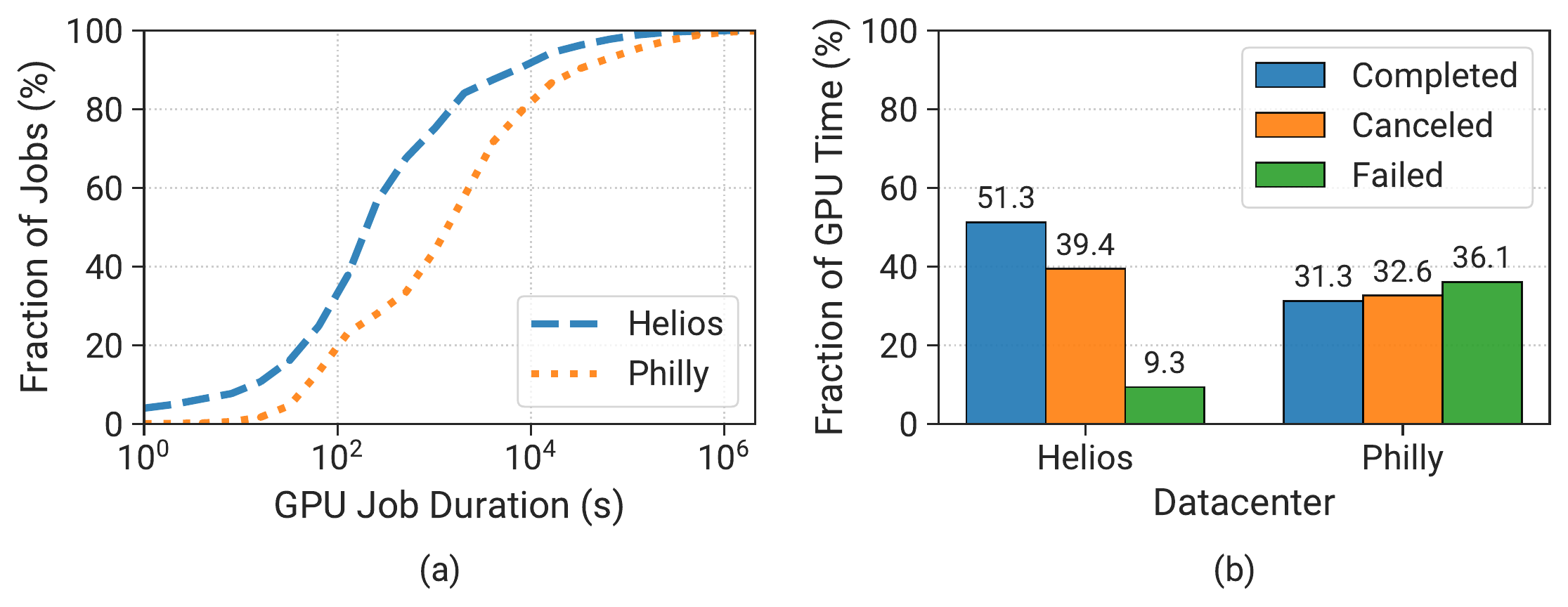}
  \caption{Comparisons of job characteristics between \DCName and Philly. (a) The CDFs of the GPU job duration. (b) Distribution of GPU time by the final job status.}
  \Description{philly}
  \label{philly}
\end{figure}

Table \ref{tab_philly} summarizes the comparisons between \DCName and Philly traces. Our traces are collected from 4 independent clusters, while the Philly trace only describes one cluster. \DCName contains 32.6 times more jobs than Philly and around half of jobs from \DCName are CPU jobs. We will reveal some interesting insights through analyzing different characteristics of CPU \& GPU jobs in \S\ref{sec_charact_helios}, which cannot be learned from Philly. Moreover, the average number of GPUs required by our GPU jobs is over twice that of Philly. The maximum number of requested GPUs from our traces is 2048, which is an order of magnitude higher than Philly. These indicate that our traces provides new features of GPU clusters and jobs, which can increase the plurality and generality in DL system research.

We calculate the average duration of GPU jobs from \DCName, which is much lower than Philly. There are three possible reasons: (1) Philly adopted \emph{A YARN} \cite{YARN} to schedule jobs, where failed jobs would be retried for a fixed number of times. Such retrials were counted into the entire job duration, statistically resulting in longer execution time. For instance, the longest job took about 60 days for a total of 3 attempts due to job failures, although the successful execution only took around 25 days. If we profile the Philly trace by regarding each attempt as an individual job, the average duration will drop to 17,398 seconds. (2) We keep all GPU jobs in \DCName to reflect the realistic characteristics of a GPU datacenter. These include debugging and testing jobs which are much shorter than normal training jobs, causing a lower average duration. (3) Our datacenter provides more computing resources. Users usually request more resources (two more GPUs on average) for training, which can also significantly accelerate the training process.

Figure \ref{philly}(a) illustrates the Cumulative Distribution Function (CDF) of job duration. We observe that Philly jobs statistically took more time than \DCName. This is consistent with our analysis from Table \ref{tab_philly}. Figure \ref{philly}(b) shows the percentages of GPU time occupied by jobs with different statuses. We can see a significant fraction of GPU time contributed to the jobs which were finally ended up with the failure or canceled status.
Particularly, over one-third of GPU time was wasted for the failed jobs in Philly and 9.3\% in \DCName.

\begin{figure}[t]
  \centering
  \includegraphics[width=\linewidth]{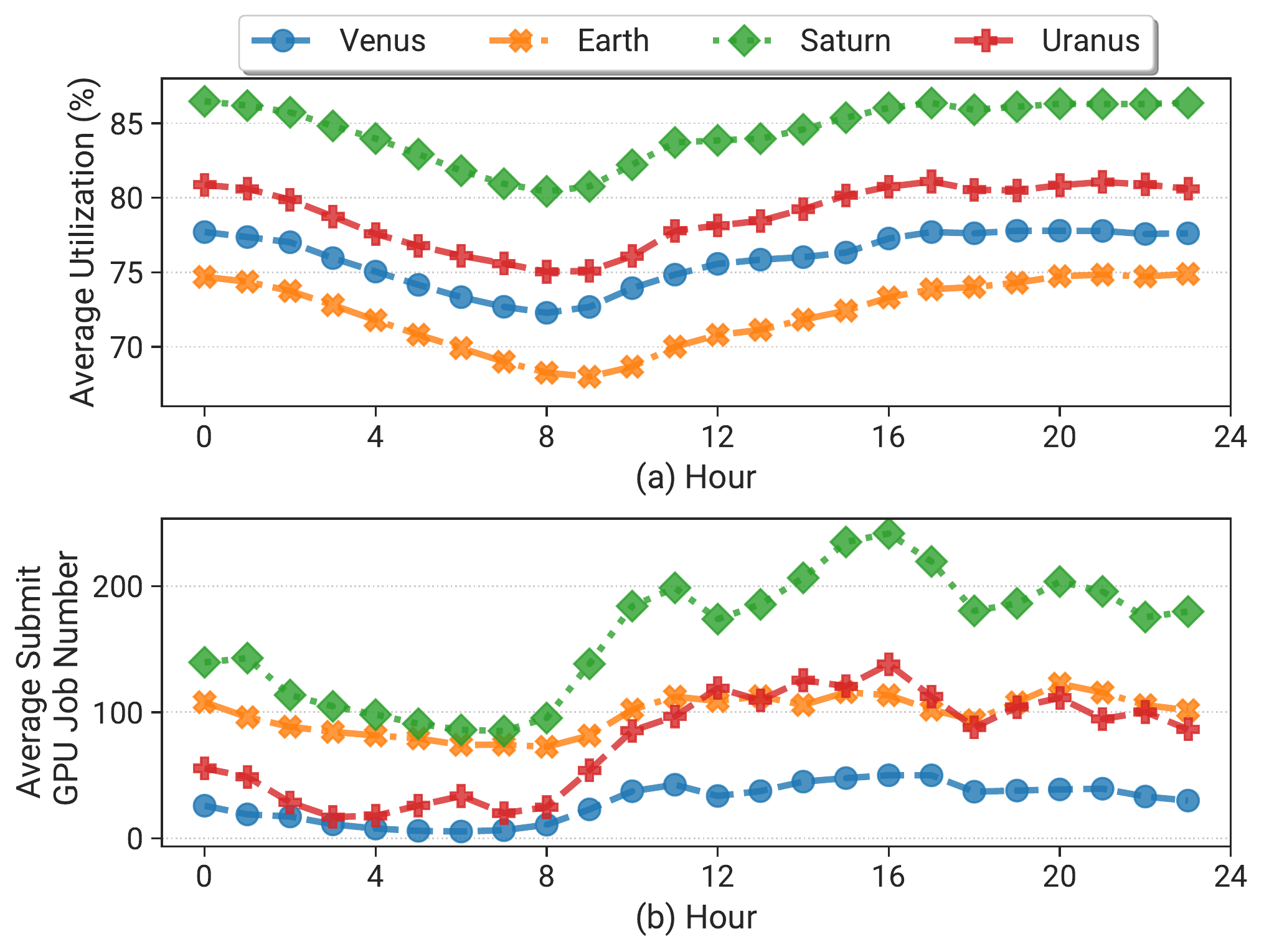}
  \caption{Daily pattern of the cluster usage in \DCName. (a) Hourly average cluster utilization over six months. (b) Hourly average GPU job submission rates over six months. }
  \Description{cluster diurnal utilization}
  \label{diurnal}
\end{figure}

\section{Characterization of DL Jobs}
\label{sec_charact_helios}
In this section, we perform a thorough analysis of our job traces. Some prior works analyzed the traditional big data traces from real-world datacenters \cite{Borg, GoogleTrace12, ServerlessATC20, BigDataTrace}. In contrast, very few studies focused on the analysis of DL jobs. \cite{Philly, AliPAI} performed an empirical study of the characteristics of their clusters. We give more comprehensive analysis from the perspectives of clusters (\S\ref{sec_charact}), jobs (\S\ref{sec_char_job}) and users (\S\ref{sec_user}).

Our traces cover different characteristics of DL jobs and behaviors of AI developers and researchers in SenseTime. For instance, users can submit long-term production jobs, as well as short-term exploratory jobs for debugging purposes. Users might early stop their jobs when they can (not) reach the expected performance. We perform our assessment statistically, and believe the conclusions are general for other organizations and clusters as well.

\subsection{Cluster Characterization}
\label{sec_charact}

\subsubsection{Daily Trends of Cluster Usage}
\label{ssec_diurnal}

Figure \ref{diurnal}(a) shows the average cluster utilization for every hour in one day. All the clusters and users are in the same timezone, and hence exhibit similar patterns for the daily cluster usage. The utilization of all the clusters ranges from 65\% to 90\%. \texttt{Saturn} has the highest utilization, while \texttt{Venus} and \texttt{Earth} are relatively underutilized. The standard deviation of hourly utilization in \texttt{Saturn} is 7\% and ranging from 10\% to 12\% in other clusters. Besides, we observe a 5\textasciitilde 8\% decrease at night (0 am -- 8 am) for all the clusters, which is not very significant. This is because the workloads in \DCName are mainly DL training jobs, which can take hours or days to complete. It is common that some jobs are submitted in the daytime but still keep running overnight.

Figure \ref{diurnal}(b) shows the average GPU job submission rate for each hour during the six months.
All the clusters have similar patterns of the job submission in one day: the number drops to the lowest point at night (sleep), and experiences a slight drop around 12pm (lunch) and 6pm (dinner). It is also interesting to note that \texttt{Earth} has a stable and high submission rate (\textasciitilde 100 jobs) per hour, but has the lowest utilization among the four clusters. This is because the GPU jobs in \texttt{Earth} are overall shorter than the other clusters. The cluster utilization depends on both the number of jobs as well as their running time.
Further, the frequency of job submission is much lower than big data clusters \cite{Borg, diversity}. This implies some time-consuming scheduling optimization algorithms would apply for scheduling DL training jobs.




\Imp{1}{Both the cluster utilization and the job submission rate exhibit obvious daily patterns. This provides us opportunities to predict those behaviors in advance and then perform the optimal resource management and job scheduling.}



\begin{figure}[t]
  \centering
  \includegraphics[width=\linewidth]{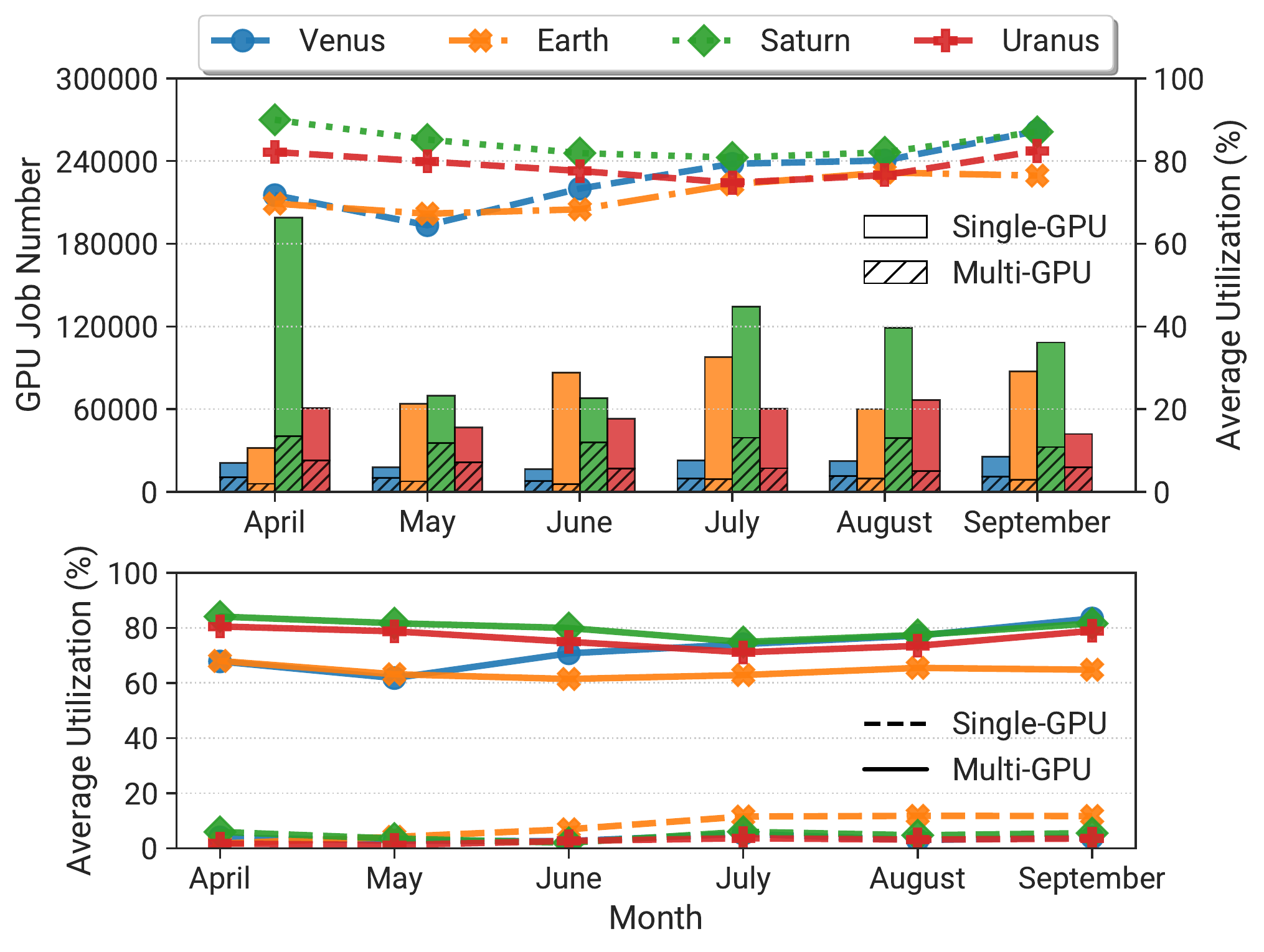}
  \caption{Monthly trends of cluster activities in \DCName. Top: number of submitted (single- and multi-GPU) jobs (\emph{bars}) and average cluster utilization (\emph{dashed lines}).  Bottom: average cluster utilization from multi-GPU jobs (\emph{solid lines}) and single-GPU jobs (\emph{dashed lines}).}
  \Description{monthly trend}
  \label{monthly}
\end{figure}


\subsubsection{Monthly Trends of Cluster Usage}
\label{ssec_monthly}
We further analyze the monthly trends of GPU resources and job behaviors. Figure \ref{monthly} (top) shows the monthly statistics of GPU job submission\footnote{Our traces end on September 27th. So the reported numbers of September are around 10\% lower than the actual one.} (bars) and average cluster utilization (dashed lines). All the clusters have stable submissions of multi-GPU jobs each month, while the numbers of single-GPU jobs fluctuate dramatically. The utilization of \texttt{Saturn} and \texttt{Earth} remains stable under varied numbers of GPU jobs per month. Surprisingly, \texttt{Saturn} executed almost twice GPU jobs in July compared with May or June, whereas the utilization in May (85.21\%) and June (81.92\%) is even higher than July (80.87\%). Furthermore, we find the distribution of multi-GPU jobs within the same cluster is similar each month.
The average number of requested GPUs is very close with a standard deviation of 2.9. Therefore, we can accurately predict the monthly submissions of multi-GPU jobs from the previous months' data. Figure \ref{monthly} (bottom) presents the cluster utilization contributed by single-GPU and multi-GPU jobs. It is evident that single-GPU jobs have little influence on the overall cluster utilization (less than 6\% except for \texttt{Earth}). Conversely, multi-GPU jobs are dominant to cluster utilization.


\Imp{2}{For monthly trends, it is infeasible and unnecessary to predict the submissions of single-GPU jobs due to their weak impact on the cluster usage. In contrast, multi-GPU jobs exhibit more stable monthly patterns, and are critical to cluster utilization, which we can predict for better scheduling efficiency.}






\begin{figure}[t]
  \centering
  \includegraphics[width=\linewidth]{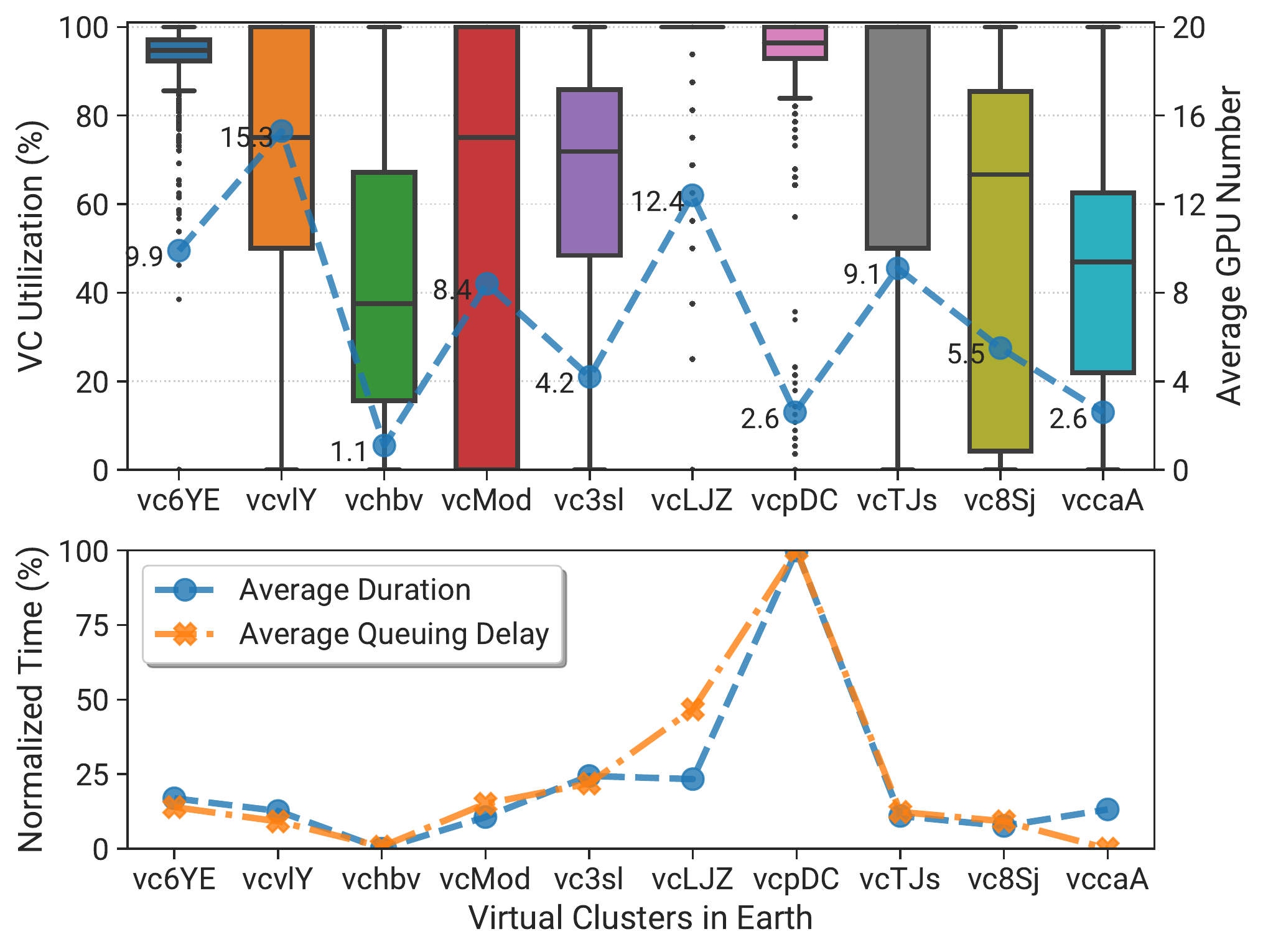}
  \caption{VC behaviors in \texttt{Earth}. Top: The boxplot of utilization distributions for the top 10 largest VCs and the job's average number of requested GPUs in each VC (\emph{dashed line}). Bottom: Min-Max normalized average job duration (blue dashed line) and queuing delay (orange dashed line).}
  \Description{box vc utilization}
  \label{box_vc_utilization}
\end{figure}

\subsubsection{Virtual Cluster Behaviors}
\label{ssec_VC}
In addition to the entire physical clusters, investigation of VCs is also indispensable. We select a period when the VC configuration remains stable (May in \texttt{Earth}). Figure \ref{box_vc_utilization} (top) shows the utilization distributions of the 10 largest VCs (in descending order) averaged per minute. Specifically, there are 208 GPUs in \textit{vc6YE} and 32\textasciitilde 96 GPUs in other VCs. Each box is framed by the first and third quartiles, while the black line inside the box represents the median value. Both whiskers are defined at 1.5 times the InterQuartile Range (IQR). We plot the job's average number of requested GPUs for each VC above the corresponding box. Figure \ref{box_vc_utilization} (bottom) shows the average queuing delay and job duration for each VC.

We find the behaviors of each VC vary significantly, as they run different types of GPU jobs in terms of resource demands and duration. First, we observe that
the VC utilization is positively correlated with the average GPU demands. \textit{vc6YE} and \textit{vcLJZ} keep over 90\% utilization most of the time as they generally run large jobs. In contrast, the utilization of \textit{vchbv} and \textit{vccaA} is basically below 65\% for hosting small jobs. One exception is \textit{vcpDC}, which has high utilization but small average numbers of GPUs. Second, the job queuing delay is approximately proportional to the average job duration. Busy VCs (e.g., \textit{vcLJZ} and \textit{vcpDC}) typically have much longer queuing delay. These prove that job queuing and resource underutilization co-exist in our clusters due to imbalanced VCs.

The key reason is the adoption of static partitioning with VCs. This simple and mature solution is widely used in production GPU clusters (e.g., Microsoft \cite{Philly, HiveD}, Alibaba \cite{AntMan, AliPAI}) for fairness among multi tenants. However, it also causes long queuing delay and resource underutilization. Researchers also designed advanced scheduling algorithms to address these issues with high fairness \cite{Themis, Gandivafair, HiveD}. How to implement them into production clusters with better reliability and robustness will be an important future work.

\Imp{3}{Different groups submit DL jobs to their VCs with distinct GPU demands and duration. Hence, the imbalanced resource allocation across VCs can lead to low resource utilization and severe job queuing delay \cite{TrestlesHPC}. It is critical to consider fairness when designing schedulers for shared clusters \cite{Themis, Gandivafair, HiveD}.} 


\subsection{Job Characterization}
\label{sec_char_job}
Beyond the cluster statistics, our traces also contain rich information at the job level. We draw interesting conclusions from such information, as discussed below.

\subsubsection{Job Execution Time}
\label{ssec_jobtime}

\begin{figure}[t]
  \centering
  \includegraphics[width=\linewidth]{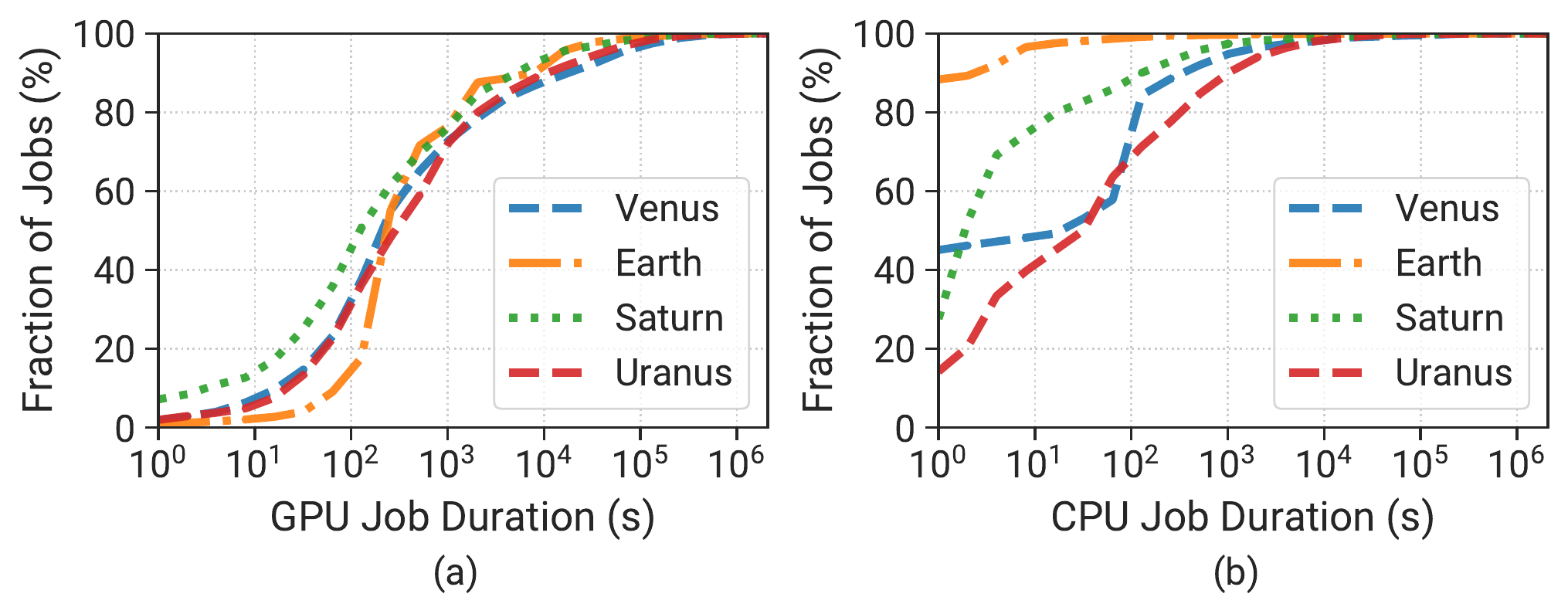}
  \caption{The CDFs of (a) GPU and (b) CPU job duration.}
  \Description{cdf job time}
  \label{cdf_job_time}
\end{figure}

As shown in Table \ref{tab_philly}, the number of GPU jobs is close to the number of CPU jobs in the \DCName traces. However, the average execution time of GPU jobs (6,652s) is 10.6$\times$ longer than CPU jobs (629s). More than 50\% of CPU jobs run for less than 2s. In contrast, the median execution time of GPU jobs is 206s. More specifically, Figure \ref{cdf_job_time} compares the duration distributions of GPU and CPU jobs in each cluster. The duration of GPU jobs ranges from seconds to weeks, and is generally over an order of magnitude longer than CPU jobs. Interestingly, these four clusters have diverse duration distributions of CPU jobs, while similar distributions for GPU jobs. In \texttt{Earth}, short-term CPU jobs account for a larger portion compared with other clusters: nearly 90\% of CPU jobs run for only one second in \texttt{Earth}. Most of them are related to training progress and node state queries. As for GPU jobs, roughly three-quarters of jobs last for less than 1000 seconds as they are mainly for model evaluation and program debugging.


\begin{figure}[t]
  \centering
  \includegraphics[width=\linewidth]{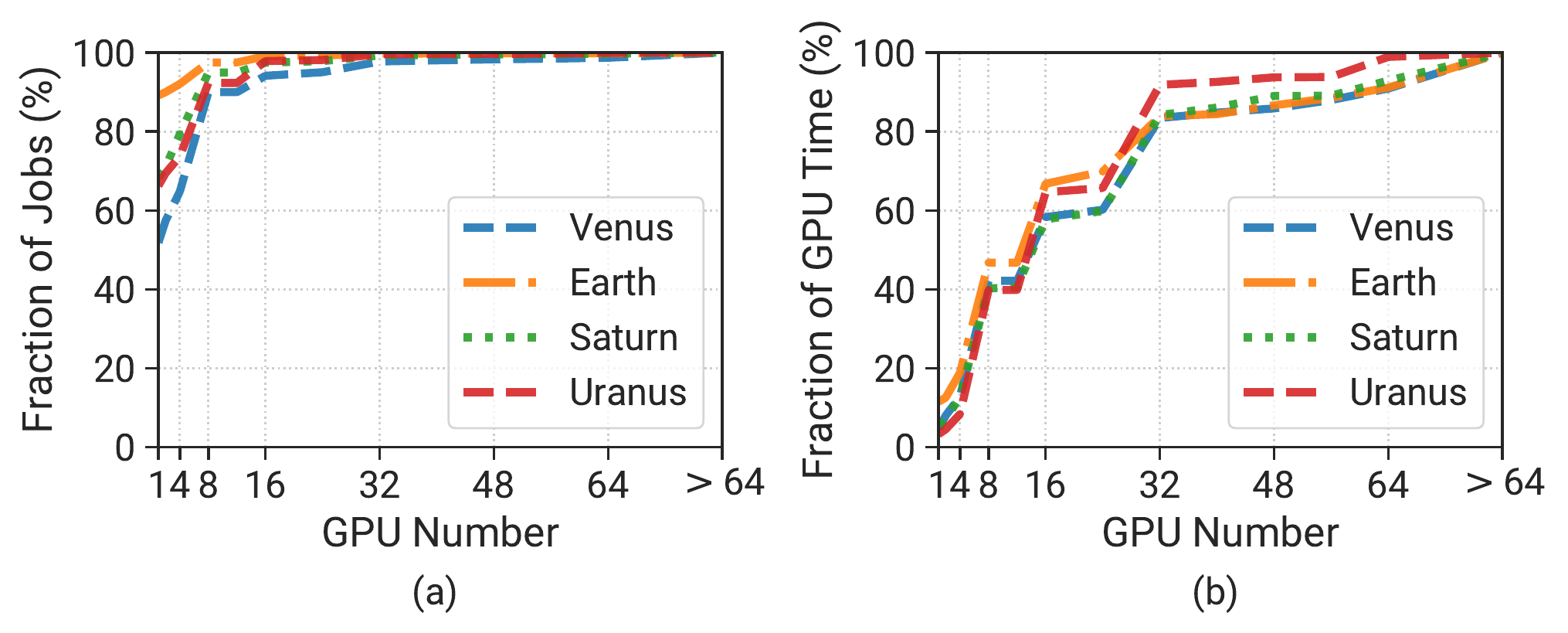}
  \caption{The CDFs of job sizes (in GPU number) with the number of (a) jobs and (b) GPU time.}
  \Description{cdf job gpu num}
  \label{cdf_job_gpu_num}
\end{figure}

To dive deeper into the characteristics of GPU jobs in each cluster, we investigate the relationship between the GPU demands, GPU time and number of GPU jobs. We show the CDFs of requested GPU demands with the number of jobs (Figure \ref{cdf_job_gpu_num}(a)) and GPU time (Figure \ref{cdf_job_gpu_num}(b)). We observe there are over 50\% single-GPU jobs in each cluster, and the largest ratio is 90\% in \texttt{Earth}. However, they only occupy 3\textasciitilde12\% of the total GPU time. In contrast, although the proportion of large-size jobs ($\geq$8 GPUs) is smaller than 10\%, they account for around 60\% of computing resources.



\Imp{4}{Despite the number of single-GPU jobs is predominant, GPU resources are mainly consumed by multi-GPU jobs. Hence, optimization of multi-GPU jobs is more important to improve cluster efficiency.
  This characteristic resembles traditional HPC workloads \cite{diversity, TraceSC20, Rodrigo2018} and implies some optimization techniques in HPC can also be applied to GPU clusters.}


\subsubsection{Job Final Statuses}
\label{ssec_status}

\begin{figure}[t]
  \centering
  \includegraphics[width=\linewidth]{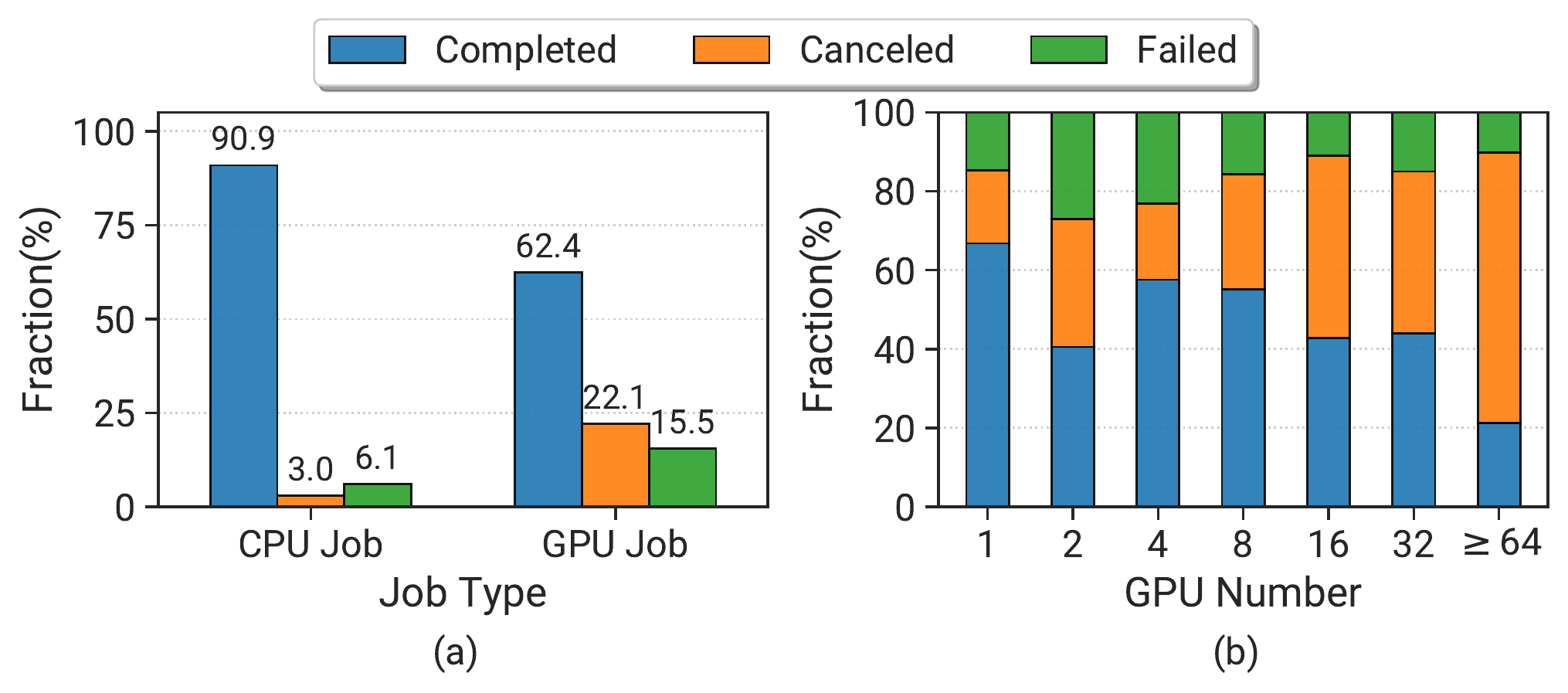}
  \caption{Distribution of jobs by their final statuses. (a) Comparisons of CPU and GPU jobs for their final statuses. (b) Percentages of final job statuses w.r.t. different GPU demands.}
  \Description{job state}
  \label{job_state}
\end{figure}

Figure \ref{job_state}(a) summarizes the distributions of final statuses for CPU and GPU jobs of these 4 clusters. The ratio of unsuccessful GPU jobs (37.6\%) is significantly higher than CPU jobs (9.1\%). One reason is that some users prefer to terminate their DL training jobs in advance as the model accuracy already converges to the satisfactory value. These canceled jobs are still successful. Another reason is that users can inspect the training states and kill the poor-performing jobs earlier. This reflects the \emph{feedback-driven exploration} feature of DL training jobs. This early-stopping feature can be leveraged to optimize the scheduling efficiency. For instance, \cite{Optimus,Gandiva} help users automatically make the early-stopping decision through recording training feedback and predicting future training performance. This can bring a huge benefit to datacenters.

There are also other causes for failed jobs, including timeout, node failure, incorrect inputs, runtime failure, etc. Microsoft \cite{Philly, DLFailures} presented a detailed analysis of the reasons for training job failures, so we do not conduct similar investigations in this paper.

\Imp{5}{Since many DL training jobs can reach the convergence earlier than expected, the scheduler can automatically detect this condition and stop the jobs for resource efficiency \cite{Optimus,Gandiva}. Users can use different metrics (e.g., loss, accuracy) and authorize the scheduler to monitor and manage their jobs.}


In Figure \ref{job_state}(b), we quantify the ratios of different job final statuses in terms of GPU demands. We only consider the GPU numbers of $2^k(k \in\mathbb{N})$ as they are mostly requested in \DCName. We observe that the ratio of job completion keeps decreasing as the number of GPUs increases, with an exception of 2-GPU jobs. For large jobs with 64 or more GPUs, only fewer than a quarter of jobs complete successfully while the canceled ratio even reaches roughly 70\%. This is because these jobs typically run for very long time, and users have higher chances to early stop them to save time and resources.

Additionally, we find
most failed jobs are terminated within a short time, which matches the conclusions in prior works \cite{Philly, DLFailures}. The majority of failures are incurred by user errors, such as script configuration, syntax/semantic errors in the program. However, plenty of short-term debugging jobs suffer from severe queuing delays. Users usually fail to get the code debugging feedback timely, which considerably affects their experience.

\Imp{6}{A lot of failed jobs are for debugging purposes, and last for a very short time. However, they are mixed with the long-term production jobs in the queue and possibly suffer from much longer waiting time than execution. A possible solution is to allocate a special VC for debugging jobs and enforce a short-term limit (e.g., 2 minutes). This can help users obtain error messages timely and filter most failed jobs for normal clusters.}



\begin{figure}[t]
  \centering
  \includegraphics[width=\linewidth]{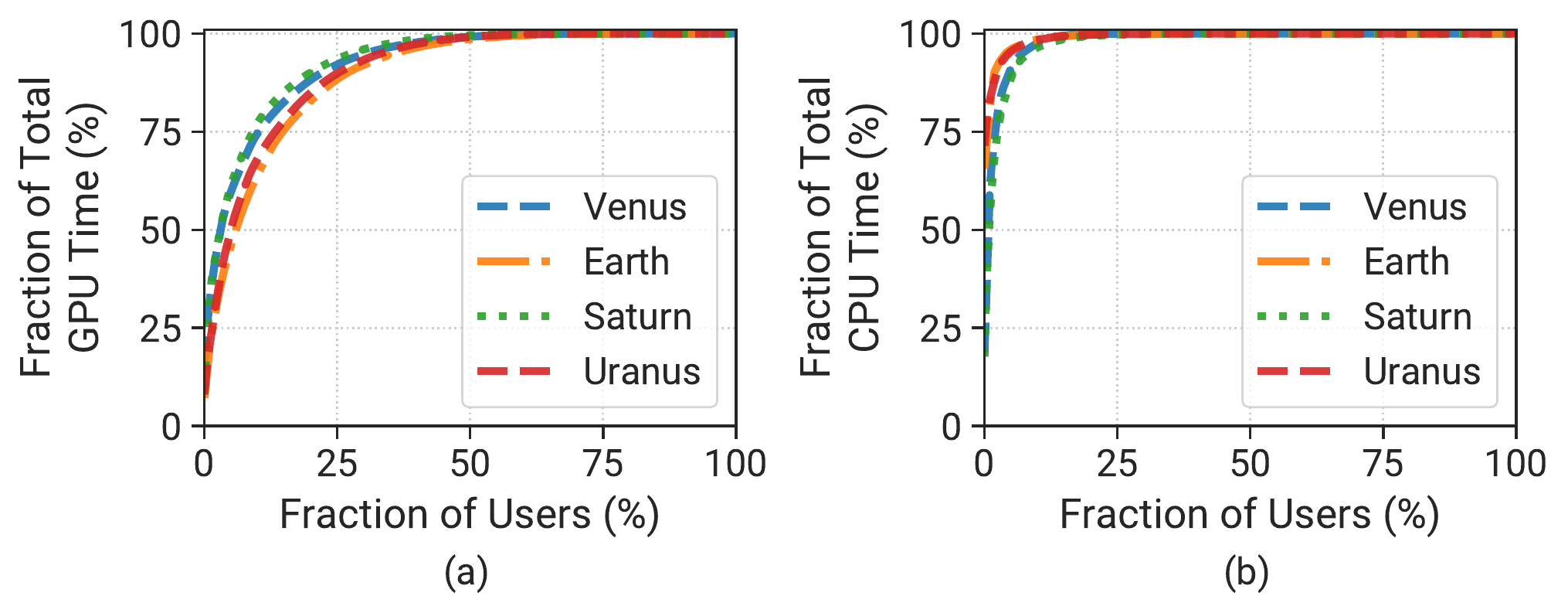}
  \caption{The CDFs of users that consume the cluster resources in terms of (a) GPU Time (b) CPU Time.}
  \Description{cdf time user}
  \label{cdf_time_user}
\end{figure}

\begin{figure}[t]
  \centering
  \includegraphics[width=\linewidth]{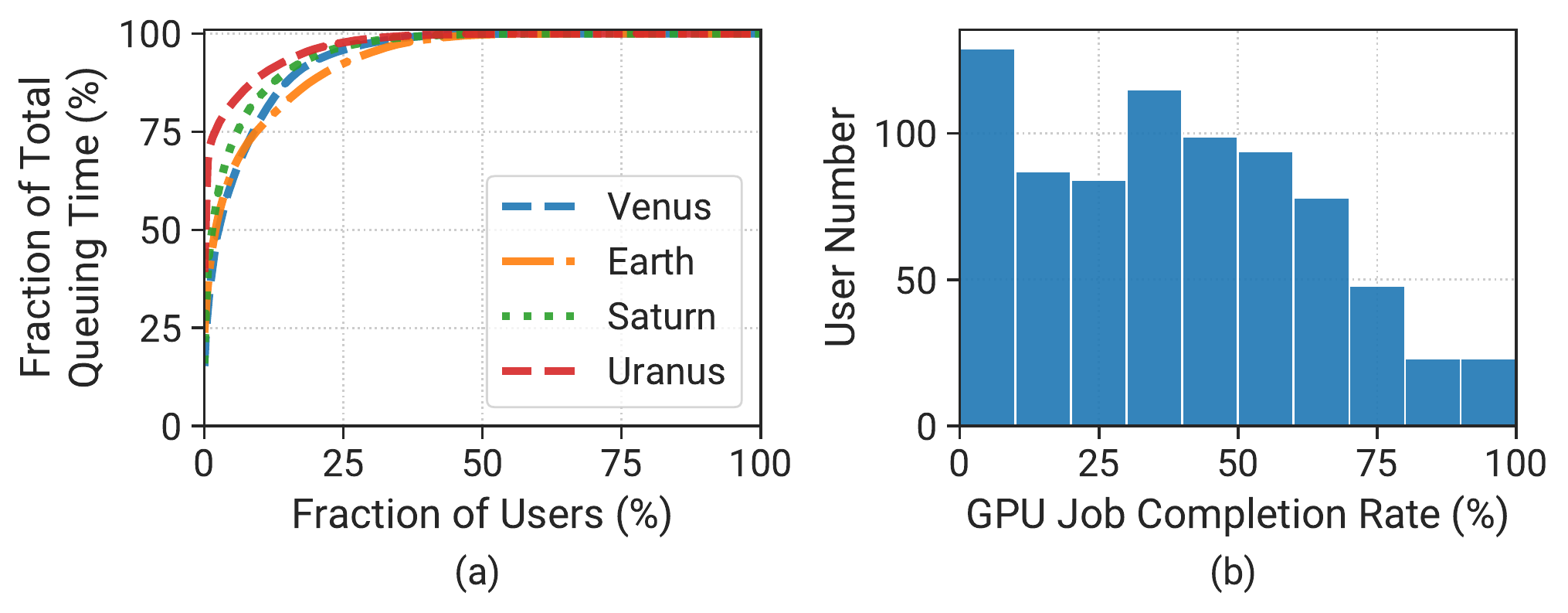}
  \caption{(a) The CDFs of users w.r.t. GPU job queuing delay. (b) The distribution of user GPU job completion ratios.}
  \Description{cdf pend user}
  \label{cdf_pend_user}
\end{figure}

\subsection{User Characterization}
\label{sec_user}
We analyze the traces from the users' perspective, to discover methods for user experience enhancement. This has never been considered in prior works about DL job analysis. Each cluster has  200\textasciitilde400 users. Some users can submit jobs to multiple clusters concurrently.

Similar to our analysis of job trends, we first investigate the consumption of CPU and GPU resources at the user level, as shown in Figure \ref{cdf_time_user}. We find the trends are similar across all the clusters. Compared with GPU time, the CDF curves of CPU time are much steeper, indicating CPU jobs are more concentrated within a small portion of users. This is because only 25\% of users on average need to conduct CPU tasks (e.g., video frames extraction), and the top 5\% of users occupy over 90\% CPU time. In contrast, almost every user has GPU training jobs, and the top 5\% of users consume 45\textasciitilde60\% GPU time.

Next, we study the distributions of GPU job queuing delay among users, as shown in Figure \ref{cdf_pend_user}(a). We observe that most users do not suffer from severe job queuing, whereas a few users have jobs blocked for a long time. In \texttt{Uranus}, the top 1\% of users (only 3) bear over 70\% queuing time, even they are not among the top 10 resource-consumption users. We name them ``marquee users'' \cite{Wolter2006}, and their experiences need to be ameliorated.

Figure \ref{cdf_pend_user}(b) shows the distributions of users for different GPU job completion rates.
It is obvious that the users' GPU job completion rates are generally low, which proves that the high fraction of unsuccessful GPU jobs (shown in Figure \ref{job_state}) reflects the users' overall behaviors instead of some individual ones.

\Imp{7}{To alleviate the problem of unfair cluster queuing, it is recommended that the scheduler should consider our user-level analysis to perform the corresponding optimization. For instance, the scheduler can dynamically adjust temporary priorities to users, especially to the marquee ones, based on their current job queuing statuses. The VC configuration can also be regulated appropriately according to users' behaviors.} 


\section{A Prediction-Based Framework}
\label{sec:system}
From \S\ref{sec_charact_helios},
we find the feasibility of predicting clusters' behaviors (e.g., job duration, node states) from the history. Inspired by this observation, we design a novel prediction-based GPU resource management framework, which leverages the historical data to improve the resource usage efficiency and workload performance.

\subsection{Framework Overview}
\begin{figure}[ht]
  \centering
  \includegraphics[width=\linewidth]{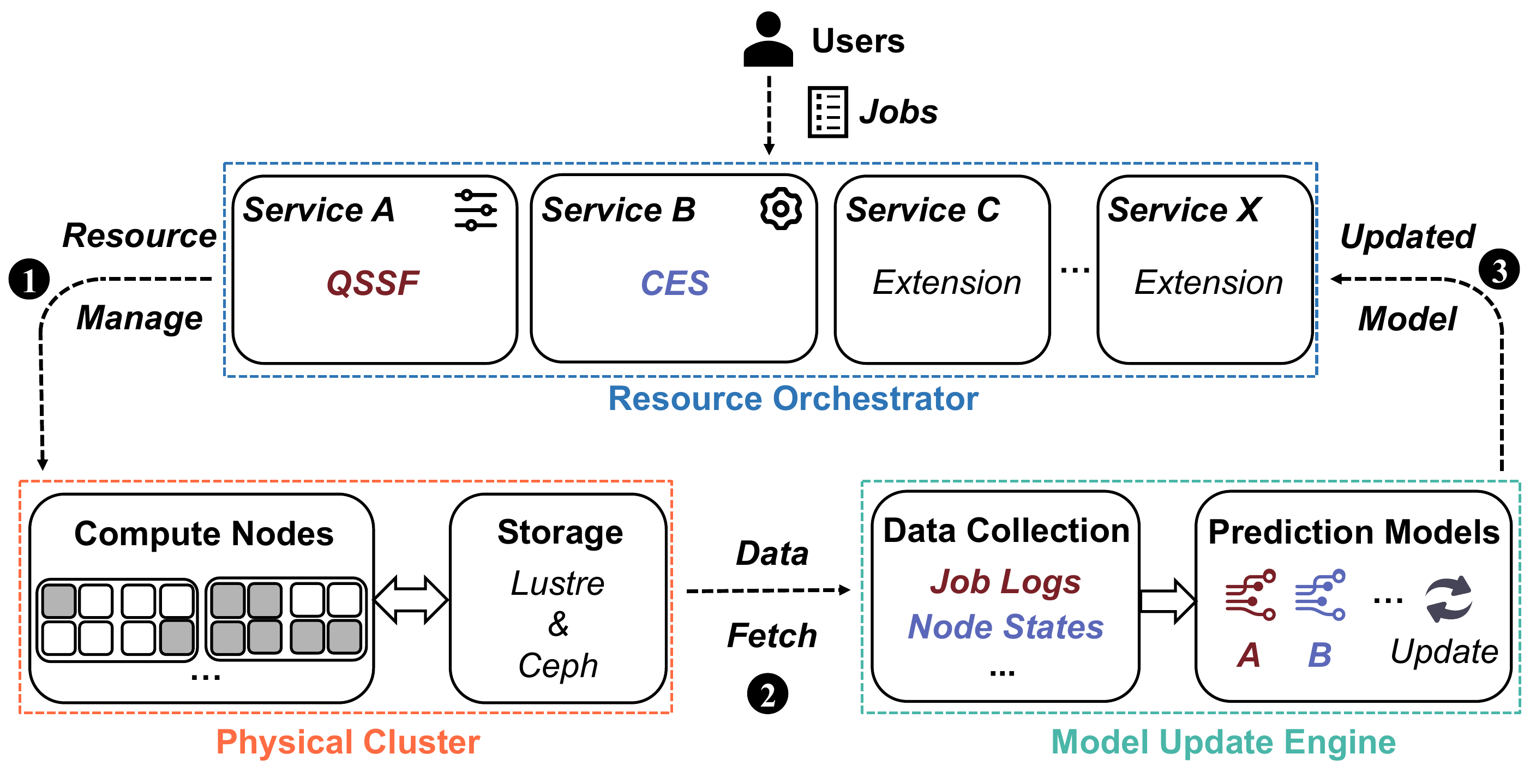}
  \caption{Overview of our prediction-based framework.}
  \Description{Framework}
  \label{Framework}
\end{figure}

Figure \ref{Framework} illustrates the overview of our framework. It is designed as a centralized manager built atop each GPU cluster. It adopts the ``plug-and-play'' fashion, where different resource management services can be integrated into this framework. Each service is independent and targets a different perspective of optimization. They share common design philosophy, and follow the same workflow. The cluster operators can select services based on their demands.

The framework consists of two components: \textit{Model Update Engine} and \textit{Resource Orchestrator}. For each service, a machine learning model is trained to predict the job behaviors or cluster states. During the operation, the \textit{Resource Orchestrator} uses the model to predict the upcoming \emph{events} and determines the optimal \emph{resource management} operation ({\tiny\encircle{\normalsize1}}), which will be executed in the cluster. Meanwhile,  the \textit{Model Update Engine} fetches the run-time data ({\tiny\encircle{\normalsize2}}) regularly (e.g., every minute) or triggered by events, and fine-tunes the model periodically to adapt to the changes in the cluster ({\tiny\encircle{\normalsize3}}).


Our framework has several benefits. (1) \emph{Extensibility}: we provide an abstract pipeline for resource management. As case studies, we design two novel services: \textit{Quasi-Shortest-Service-First Scheduling} to minimize the cluster-wide average JCT (\S\ref{ssec_QSSF}), and \textit{Cluster Energy Saving} to improve the cluster energy efficiency (\S\ref{ssec_CES}). Other services based on machine learning prediction can also be integrated into our framework, e.g., burstiness-aware resource manager \cite{Burstiness16, Burstiness17}, network-aware job scheduler \cite{NetworkAware, HIRE}, etc. (2) \emph{High usability}: our framework can be deployed into arbitrary GPU clusters. The services work as plugins atop the current management systems (e.g. \emph{Slurm}, \emph{YARN}, \emph{Kube-scheduler}) with minimal or no modifications to them. Users do not need to provide extra information or specifications. (3) \emph{Low overhead}: the prediction and operation latency for each service typically takes milliseconds, which is negligible for DL workloads.

\begin{figure*}[t]
  \centering
  \includegraphics[width=\linewidth]{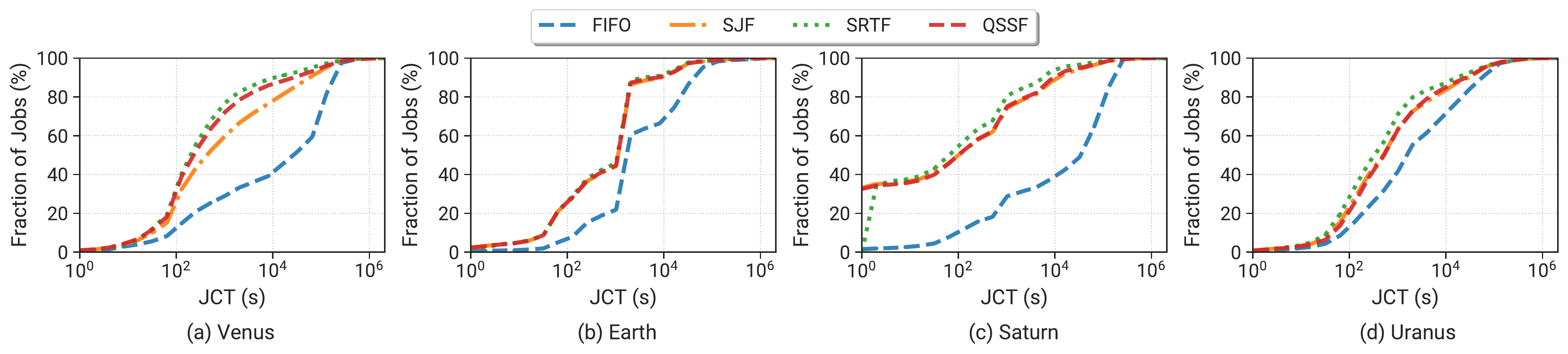}
  \caption{Comparisons of JCT distributions using different scheduling algorithms, based on the September job traces across 4 clusters in \DCName. Note that SJF and SRTF are optimal baselines (with perfect job duration information).}
  \Description{simu jct}
  \label{simu_jct}
\end{figure*}

\subsection{Quasi-Shortest-Service-First Scheduling}
\label{ssec_QSSF}

\subsubsection{Motivation.}
All GPU clusters in \DCName are managed by \emph{Slurm}. Jobs are scheduled using the vanilla First-In-First-Out (FIFO) algorithm, similar to \emph{YARN}’s Capacity Scheduler \cite{YARN}. Due to such runtime-unaware scheduling style \cite{Tumanov2016}, users complain that even short-term jobs can suffer from long queuing delays.

Both Shortest-Job-First (SJF) and Shortest-Remaining-Time-First (SRTF) algorithms were proposed
to reduce the average JCT \cite{Altruistic, GRAPHENE, Tiresias, E-LAS} with and without preemption respectively. However, they are too ideal and thus impractical in GPU clusters, due to the following two reasons: (1) some jobs in GPU clusters do not have iterative characteristics, and cannot be preempted and then restored from the checkpoints. Hence, the preemption-enabled scheduling algorithms (SRTF) are not applicable for these jobs.
(2) These algorithms rely on the information of job duration or remaining execution time, which is uncertain and could be affected by many unexpected factors (e.g., termination in advance, training early-stopping \cite{Prechelt1998}, error and node crash, etc.).
To address this issue, prior works try to obtain the job runtime information from the users \cite{BorgK8S, Mesos, YARN}, job profiling  \cite{Dryad,Jockey,ARIA,Apollo, Ernest} or leveraging the job's periodic features \cite{ReservationSched, NetworkAware, Hawk, Morpheus}. These approaches can either affect the usability or operation cost, or lack of generality for different types of workloads.

Driven by these limitations, we design a new Quasi-Shortest-Service-First (QSSF) scheduling service. This scheduler adopts the non-preemption mechanism, which can be applied to different types of jobs in our datacenter. Owing to the constraint of the gang scheduling for DL training jobs, large-size short-term jobs can occupy many GPUs, which can block multiple small-size jobs using the SJF scheduler \cite{Tiresias}. We choose to rank jobs with GPU time instead of duration. Through the trace analysis, we find there exist strong correlations between the job duration and some attributes, such as job name, user, GPU demands, submission time, etc. Hence, our scheduler leverages these attributes to predict the jobs' priority orders (i.e., expected GPU time) for scheduling. Similar ideas were also applied for scheduling big data workloads \cite{Tumanov2016, 3Sigma}. We consider more attributes with a machine learning model ensemble to enhance the scheduling performance for DL jobs.

\begin{algorithm}[t]
  \caption{Quasi-Shortest-Service-First Scheduler}
  \small
  \label{alg1}
  \begin{algorithmic}[1]
    \Input \textbf{New job:} $\mathcal{J}$, \textbf{VCQueue:} $\mathbb{Q}$, \textbf{Historical job trace:} $\mathbb{J}$
    \Procedure{QSSF Schedule}{$\mathcal{J}$, $\mathbb{Q}$, $\mathbb{J}$}
    \State $\mathcal{J}.priority$ = Priority($\mathcal{J}$, $\mathbb{J}$) \Comment{Assign priorty to the new job}
    \State Enqueue $\mathcal{J}$ to $\mathbb{Q}$
    \State SortJobPriority($\mathbb{Q}$) \Comment{Sort by job priority}
    \For{all \textsl{Job} $\in \mathbb{Q}$}
    \If {Consolidate(\textsl{Job}) is \textbf{True}}
    \State ConsolidateAllocate(\textsl{Job}) \Comment{Job placement}
    \State Dequeue \textsl{Job} from $\mathbb{Q}$
    \Else
    \State \textbf{break}
    \EndIf
    \EndFor
    \EndProcedure

    \State

    \Function{Priority}{$\mathcal{J}$, $\mathbb{J}$}
    \label{func1}
    \If {UserMatch($\mathcal{J}.user, \mathbb{J}.users$) is $\varnothing$}
    \State $\mathcal{P}_R$ = Distribution($\mathbb{J}.durations$) \Comment{New user}
    \label{case1}
    \ElsIf {SimilarName($\mathcal{J}.name$, $\mathbb{J}.user.names$) is $\varnothing$}
    \State $\mathcal{P}_R$ = Distribution($\mathbb{J}.user.durations$) \Comment{New job name}
    \label{case2}
    \Else
    \State $\mathcal{P}_R$ = RollingEstimator($\mathcal{J}$, $\mathbb{J}$)
    \label{case3}
    \EndIf
    \State $\mathcal{P}_M$ = MLEstimator($\mathcal{J}$)
    \label{GBDT}
    \State $\mathcal{P}$ = $\mathcal{N} (\lambda \mathcal{P}_R + (1-\lambda) \mathcal{P}_M)$  \Comment{$\mathcal{N}$: GPU number of $\mathcal{J},\ $ $\lambda$: Merging coefficient}
    \label{priority}
    \State
    \Return $\mathcal{P}$
    \EndFunction
  \end{algorithmic}
\end{algorithm}

\subsubsection{Service Design}


Our QSSF scheduler builds a machine learning model to predict the expected GPU time of incoming jobs. When a user submits a DL job to the cluster, the scheduler immediately retrieves the relevant attributes (e.g., GPU \& CPU demands, job name, user id, target VC, etc) and infers the job's expected GPU time from the model. Then the scheduler selects and allocates jobs based on the predicted GPU time and cluster resource states. After the job termination, the job's final information will be collected by the \emph{Model Update Engine} for fine-tuning the prediction model.

We train a Gradient Boosting Decision Tree (GBDT) \cite{LightGBM} model to capture the overall trend of the relevant jobs. Specifically, we extract all features and actual duration from the traces to construct a training and validation set. We encode all the category features (e.g., user name, VC name, job name). For the extremely sparse and high-dimensional features of job names, we utilize the Levenshtein distance \cite{StringMatching} to cluster the names and bucketize similar ones, which convert them into relatively dense numerical values. For the time-related features (e.g., job submission time), we parse them into several time attributes, such as month, day of the week, hour, minute. Finally, we train a GBDT model that can map these job attributes to the corresponding duration.



Algorithm \ref{alg1} shows the pseudo-code of our QSSF algorithm. The core function is \textsc{Priority} (line \ref{func1}), which returns a priority value ($\mathcal{P}$) for a given job ($\mathcal{J}$). $\mathcal{P}$ is calculated as the weighted sum of a rolling estimate $\mathcal{P}_R$ and machine learning estimate $\mathcal{P}_M$. The rolling estimate $\mathcal{P}_R$ is computed directly from the historical jobs with similar attributes. There are three cases for calculating $\mathcal{P}_R$:
(1) if the job user cannot be found in the traces (new user), then $\mathcal{P}_R$ is the average duration of all the jobs with the same GPU demands in the traces (line \ref{case1}).
(2) If the traces have the records of this user, then we leverage the Levenshtein distance to find historical jobs from this user, which have similar names or formats as the incoming one. If no such historical jobs are found, then $\mathcal{P}_R$ is the average duration of all this user's jobs with the same GPU demands in the traces (line \ref{case2}).
(3) Otherwise, we compute $\mathcal{P}_R$ via exponentially weighted decay of duration of historical jobs with matched names (line \ref{case3}).
The machine learning estimate $\mathcal{P}_M$ is computed by the GBDT model (line \ref{GBDT}), which considers the overall trend of the relevant jobs.
Finally, we combine the two estimates $\mathcal{P}_R$ and $\mathcal{P}_M$, and multiply the requested GPU number ($\mathcal{N}$) to get the job's expected GPU time $\mathcal{P}$ as the priority value (line \ref{priority}), which can reflect both spatial and temporal aspects of the job \cite{Tiresias}.

\begin{figure}[t]
  \centering
  \includegraphics[width=\linewidth]{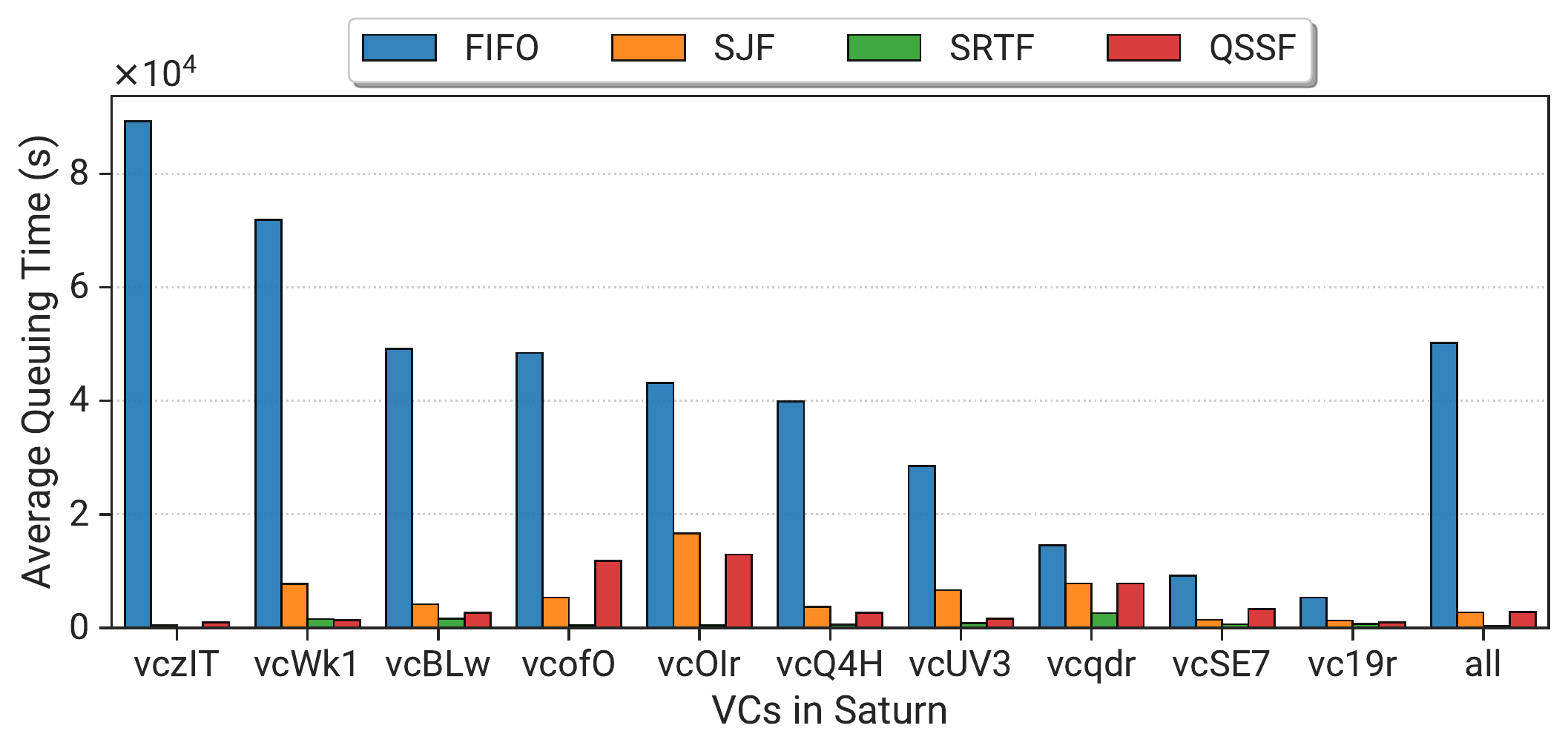}
  \caption{The average job queuing delay of the top 10 VCs in Saturn (September) with different scheduling algorithms. The column \texttt{all} represents the whole cluster.}
  \Description{simu pend}
  \label{simu_pend}
\end{figure}

We select a job from the VC queue with the highest priority (lowest predicted GPU time) for scheduling. For job placement, unless the topology is specified by the user, the \emph{ConsolidateAllocate} policy is adopted to allocate each job on as few nodes as possible to reduce the communication overhead. For instance, a 16-GPU job needs to wait for two compute nodes with 8 idle GPUs. Other placement solutions will violate the consolidation principle.

\subsubsection{Evaluation}
\label{sec:qssf-eval}
We develop a trace-driven simulator to evaluate our QSSF algorithm. It emulates a datacenter with the same configuration as \DCName in Table \ref{tab_trace}, which operates with the real-world job workflow: job arrival -- queuing -- running -- completion/canceled/failed.
Since the GPU resources are the bottleneck in our clusters, we mainly consider the GPU jobs in our simulation. We train the GBDT model using the jobs from April to August in the traces, and evaluate the model with the jobs in September.

We consider three baseline algorithms for comparisons.
(1) \emph{FIFO}: the current scheduling policy in our clusters, which is simple but has poor performance.
(2) \emph{SJF}: the optimal policy to minimize the average JCT without preemption.
(3) \emph{SRTF}: the optimal preemption-enabled version of \emph{SJF}, where we assume all the jobs can be preempted with negligible overhead. Both the SJF and SRTF algorithms are not practical, since they need perfect job duration information \cite{Tiresias, E-LAS}, which cannot be obtained in reality. They serve as the upper bound of the scheduling performance. We assume the scheduler knows the exact job duration given in the trace.
Also, we do not consider the backfill mechanism, as we want to explore how much benefit can be obtained solely from prediction. Integration of backfill with our QSSF service will be considered as future work.


\begin{figure}[t]
  \centering
  \includegraphics[width=\linewidth]{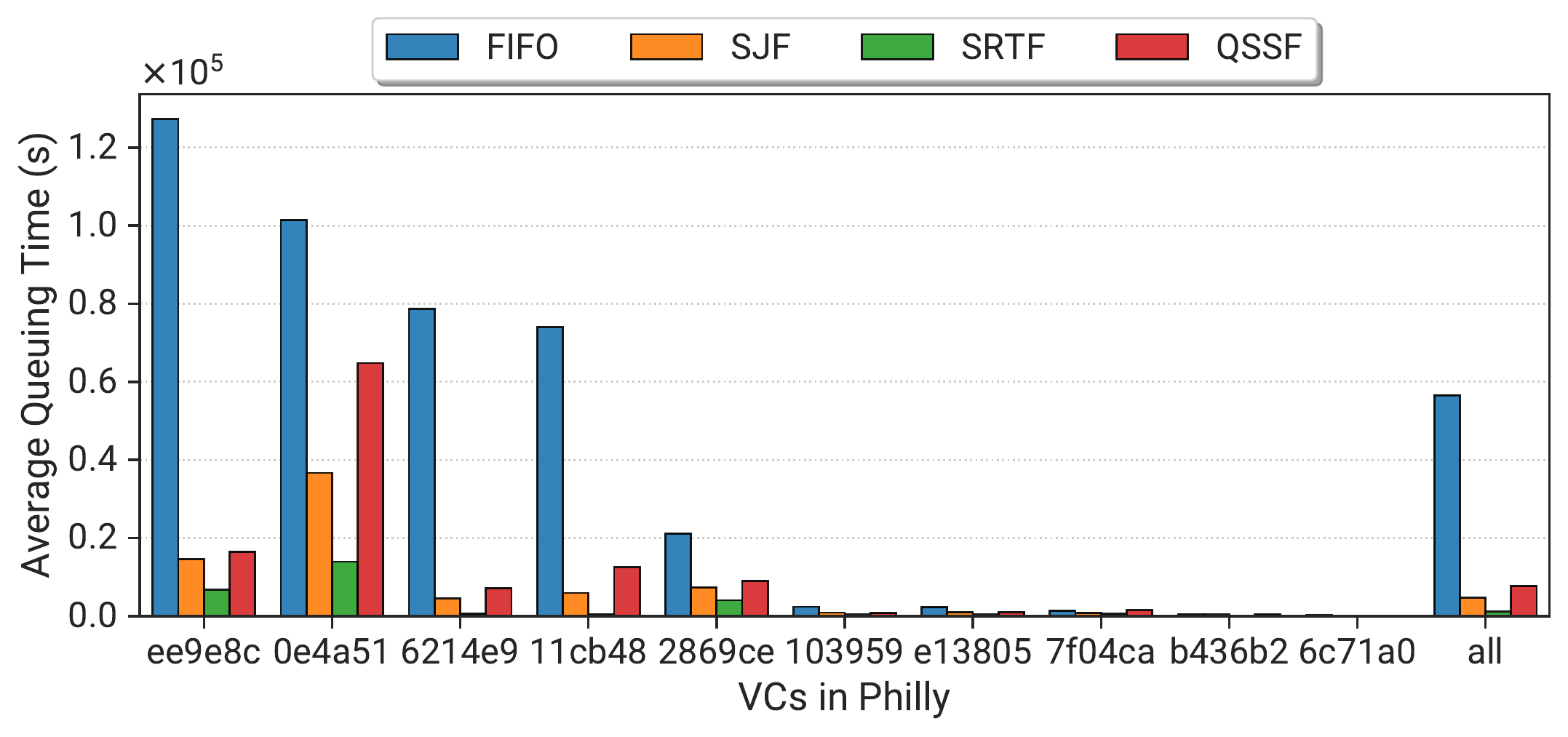}
  \caption{The average job queuing delay of the top 10 VCs in Philly (October and November) with different scheduling algorithms. The column \texttt{all} represents the whole cluster.}
  \Description{philly_pend}
  \label{philly_pend}
\end{figure}

Figure \ref{simu_jct} shows CDF curves of JCT in each cluster with different scheduling algorithms. We observe that our QSSF algorithm can significantly outperform the naive FIFO policy, and perform comparably with SRTF and SJF, without making unrealistic assumptions. The advantage of QSSF over FIFO is relatively smaller in \texttt{Uranus}, as the job queuing delay in this cluster is not as severe as the other three clusters: the queuing period takes 70\textasciitilde90\% of the total job completion time in other clusters while 42\% in \texttt{Uranus} (Table \ref{tab_qssf}).

From Table \ref{tab_qssf}, QSSF reduces 37\textasciitilde82\% of queued jobs, which is even better than SJF. Interestingly, even though QSSF prioritizes short-term jobs to alleviate the head-blocking problem, the queuing delays of large jobs are also improved because of fewer queuing jobs. Table \ref{tab_jobgroup} shows the improvement of QSSF over SJF in different groups of jobs. Short-term jobs achieve at least 9.2$\times$ improvement while long-term jobs can also obtain 2.0\textasciitilde4.8$\times$ improvement in \DCName. This justifies that QSSF will not sacrifice the interest of long-term jobs and all kinds of jobs can benefit from our service.


To evaluate the performance differences caused by the scheduler in each VC, we depict the average job queuing delay of VCs in \texttt{Saturn}, as shown in Figure \ref{simu_pend}. We select the top-10 VCs with the highest average queuing time, while the other VCs have little delay.
We observe the average queuing time of QSSF is almost identical to SJF, and remain stable in each VC. Furthermore, QSSF even outperforms SRTF in \textit{vcWk1}. This confirms the importance of GPU demands to enhance the scheduling efficiency, since large-size short-term jobs may block many small-size jobs. To summarize, compared with FIFO, QSSF achieves 1.5\textasciitilde6.5$\times$ improvement in average JCT, and 4.8\textasciitilde20.2$\times$ improvement in average queuing delay among 4 clusters in \DCName.

In addition, we also evaluate the applicability of our QSSF service on the Philly trace. Since Microsoft did not release sufficient trace information (e.g., job names and VC configurations) which is necessary for our service, we make two reasonable assumptions. (1). The VC configurations are static during our evaluation period (from 1st October to 30th November, 2017) and the size of each VC is set corresponding to its workloads. (2). The priority values are generated randomly with a similar error distribution as \DCName estimation. As shown in Table \ref{tab_qssf}, QSSF obtains comparable performance with the optimal SJF in Philly, even without precise estimates. It achieves 2.3$\times$ improvement in average JCT, 7.3$\times$ improvement in average queuing delay, and reduces 27\% of queued jobs. Figure \ref{philly_pend} presents the VC-level analysis of QSSF performance. We observe QSSF brings large improvement in each VC with regard to the average queuing time.


\begin{table}[t]
  \caption{Performance comparison of different schedulers.}
  \label{tab_qssf}
  \resizebox{1\linewidth}{!}{
    \begin{tabular}{@{}ccrrrr|r@{}}
      \toprule
      \textbf{}                                       &      & \textbf{Venus} & \textbf{Earth} & \textbf{Saturn} & \textbf{Uranus} & \textbf{Philly} \\ \midrule
      \multirow{3}{*}{\begin{tabular}[c]{@{}c@{}}Average\\ JCT (s)\end{tabular}} & FIFO & 64,702         & 19,754         & 55,984          & 19,758          & 86,072          \\
                                                      & SJF  & 21,095         & 6,892          & 8,501           & 13,226          & 34,272          \\
                                                      & QSSF & 18,349         & 6,732          & 8,581           & 13,123          & 37,324          \\\midrule
      \multirow{3}{*}{\begin{tabular}[c]{@{}c@{}}Average\\ Queuing Time (s)\end{tabular}} & FIFO & 52,933         & 13,699         & 50,202          & 8,394           & 56,531          \\
                                                      & SJF  & 9,325          & 837            & 2,719           & 1,861           & 4,731           \\
                                                      & QSSF & 6,580          & 677            & 2,798           & 1,759           & 7,783           \\\midrule
      \multirow{3}{*}{\begin{tabular}[c]{@{}c@{}}\# of \\ Queuing Jobs \end{tabular}} & FIFO & 15,336         & 30,030         & 65,991          & 16,917          & 30,282          \\
                                                      & SJF  & 8,353          & 8,848          & 21,808          & 11,320          & 21,437          \\
                                                      & QSSF & 3,713          & 5,462          & 15,311          & 10,581          & 22,026          \\  \bottomrule
    \end{tabular}}
\end{table}

\begin{table}[t]
  \caption{The ratio of queuing delay between FIFO and QSSF in different job groups. A higher ratio indicates shorter delay and better efficiency in QSSF.}
  \label{tab_jobgroup}
  \resizebox{1\linewidth}{!}{
    \begin{tabular}{@{}ccccc|c@{}}
      \toprule
                                                  & \textbf{Venus} & \textbf{Earth} & \textbf{Saturn} & \textbf{Uranus} & \textbf{Philly} \\ \midrule
      short-term (<15 mins)                       & 11.44          & 33.51          & 22.88           & 9.24            & 26.95           \\
      middle-term (15 mins\textasciitilde6 hours) & 4.13           & 13.39          & 7.39            & 2.49            & 5.54            \\
      long-term (>6 hours)                        & 3.22           & 4.77           & 3.56            & 2.00            & 1.70            \\ \bottomrule
    \end{tabular}}
\end{table}

\subsection{Cluster Energy Saving}
\label{ssec_CES}

\subsubsection{Motivation}
Electricity dominates the operation cost of modern GPU datacenters, even surpassing the manufacturing cost during the datacenters' lifetime \cite{EnergyMei}. How to reduce the energy consumption in GPU clusters becomes an important research direction. In reality, tremendous energy is wasted on idle compute nodes. Then the critical goal is to reduce energy consumption from those idle nodes while satisfying users' demands. There are generally two techniques to conserve energy in datacenters. (1) Dynamic Voltage and Frequency Scaling (DVFS) adjusts the CPU \& GPU voltage and frequency to save power. (2) Dynamic Resource Sleep (DRS) puts idle servers into deep sleep states (or simply turning them off) \cite{PowerNap, EnergyMei}. \emph{Slurm} also provides an integrated power saving mechanism \cite{SLURM} to switch the nodes between power saving and normal operation modes based on their workloads.

We introduce a novel Cluster Energy Saving (CES) service to achieve energy conservation in GPU datacenters. Existing DRS-based technique \cite{EnergyMei} simply turns off and on the nodes based on \textit{recent} and \textit{current} workloads. However, frequent server boot-up can introduce extra energy overhead and delay. In order to reduce the unnecessary mode switch operations, we build a prediction model to estimate the \textit{future} utilization trend in our clusters based on the historical node state logs. With the prediction, we can select and power off the optimal number of servers. This saves energy consumption while maintaining the cluster usability. As described in \S\ref{sec_charact}, the average utilization rate of our clusters ranges from 65\% to 90\%, and partial VCs are underutilized all the time. Hence, this strategy can bring huge financial gains for the datacenter.

\subsubsection{Service Design}
The core of our service is the prediction model, which performs a time-series forecasting task. The biggest challenge of building this model is the lack of accurate information about the future behaviors of the cluster. Therefore, we extract time-related data such as trend, seasonality, and any deterministic event as the features for prediction. Specifically, we encode repetitive patterns (e.g., hour, day of the week, date) of running nodes to explore the periodic variations. Node trends are calculated as the average values and standard deviations of active nodes under different rolling window sizes. Moreover, binary holiday indicators and various time scale lags are also crucial for prediction. We use these features to build a model which can predict the number of running nodes in the future. We try different machine learning algorithms, and find the GBDT \cite{LightGBM} model performs the best over other classical or deep learning models, e.g., ARIMA \cite{ARIMA}, Prophet \cite{Prophet}, and LSTM \cite{LSTM}. So we choose GBDT in our CES service, which can achieve around 3.6\% error rate (measured in Symmetric Mean Absolute Percentage Error (SMAPE) \cite{forecastmeasures}) in the \texttt{Earth} cluster. This can give reliable and accurate advice for powering off nodes.

\begin{algorithm}[t]
  \caption{Cluster Energy Saving Node Control}
  \label{alg2}
  \small
  \begin{algorithmic}[1]
    \Input Nodes \#: Current Active $\mathbf{C_A}$,  Current Running $\mathbf{C_R}$, Request $\mathbf{R}$, History Series $\mathcal{H}$, Prediction  Series $\mathcal{P}$
    \Procedure{JobArrivalCheck}{$\mathbf{C_A}$, $\mathbf{R}$}
    \label{wake check}
    \If {$\mathbf{C_A} < \mathbf{R}$} \Comment{Lack nodes}
    \State NodesWakeUp($\mathbf{R} - \mathbf{C_A} + \sigma$) \Comment{$\sigma$: Buffer nodes}
    \EndIf
    \EndProcedure

    \State

    \Procedure{PeriodicCheck}{$\mathcal{H}$, $\mathbf{C_A}$, $\mathbf{C_R}$, $\mathcal{P}$}
    \label{sleep check}
    \State $\mathcal{T}_H$ = RecentNodesTrend($\mathcal{H}$)
    \State $\mathcal{T}_P$ = FutureNodesTrend($\mathcal{P}$)
    \If {$\mathcal{T}_H \geq \xi_H$ \textbf{and} $\mathcal{T}_P \geq \xi_P$} \Comment{$\xi_H, \xi_P$: Thresholds}
    \State $\mathbf{C_A} \gets $ DynamicResourceSleep($\mathbf{C_R}+\sigma$ )
    \EndIf
    \EndProcedure
  \end{algorithmic}
\end{algorithm}

\begin{figure}[t]
  \centering
  \includegraphics[width=\linewidth]{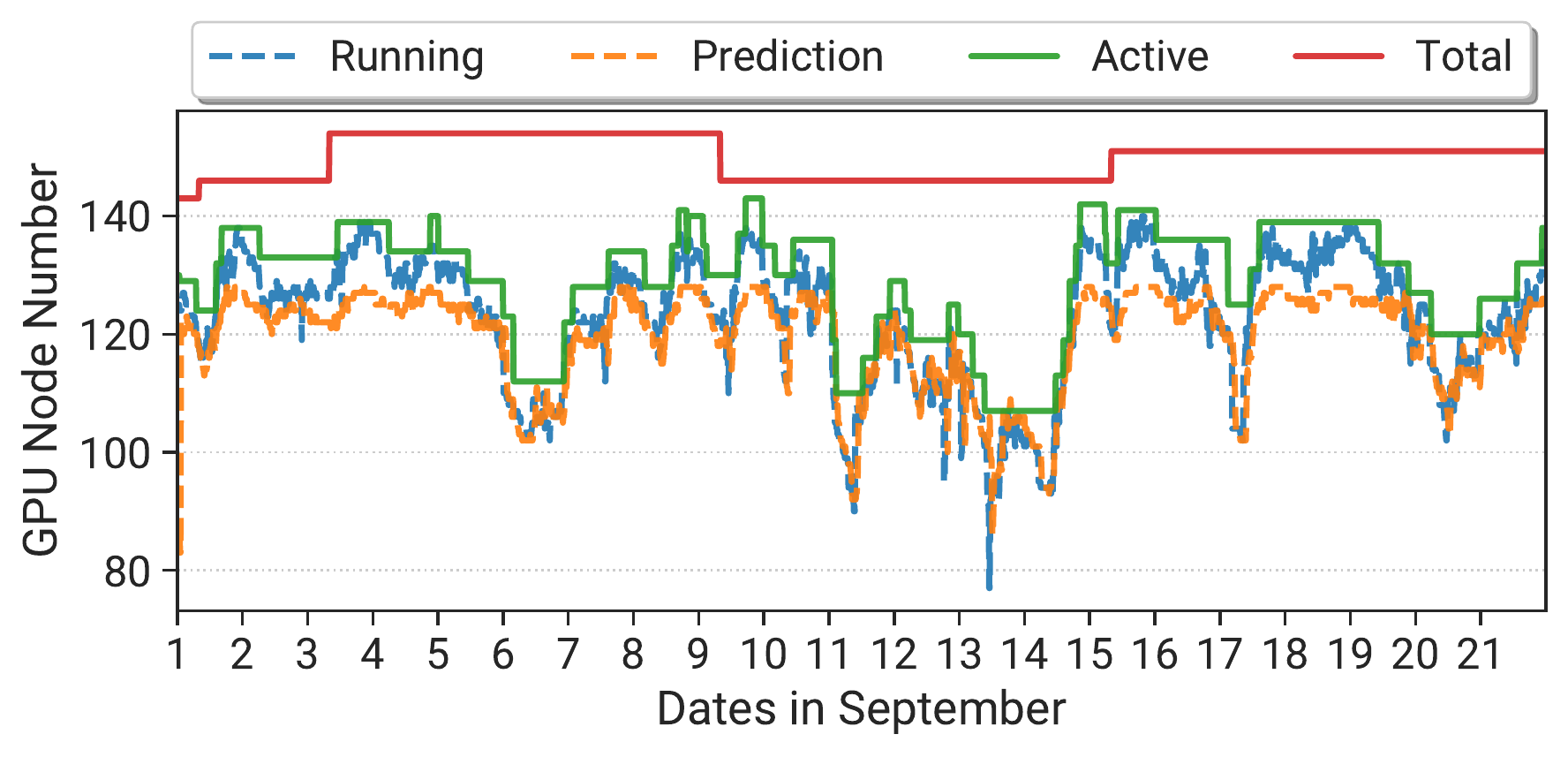}
  \caption{Numbers of compute nodes at different states in \texttt{Earth} from 1st September to 21st September (3 weeks).}
  \Description{simu energy}
  \label{simu_energy}
\end{figure}

With the prediction results, we can select the idle nodes and apply DRS to them. Algorithm \ref{alg2} illustrates the procedure, where two steps are conducted to guarantee compute resources and conserve energy.
(1) When a new job is submitted to the cluster, the service performs \textsc{JobArrivalCheck} (line \ref{wake check}) to inspect whether the requested resources ($\mathbf{R}$) surpass the currently available ones ($\mathbf{C_A}$) in the cluster. If so, the service needs to wake up some nodes immediately via the Intelligent Platform Management Interface (IPMI). The quantity of nodes is determined by the resource gap ($\mathbf{R}-\mathbf{C_A}$) with an additional number $\sigma$ to buffer unexpected future jobs.  (2) Our service also calls \textsc{PeriodicCheck} regularly (e.g., every 10 minutes) to check if extra nodes need to be powered off. The decision is made from both the historical and predicted future trends, to circumvent the incorrect DRS operations caused by the prediction error. Specifically, we call the function \emph{RecentNodesTrend} to calculate the reduced number of active nodes during a fixed past period (e.g., one hour) from the historical data. We also call \emph{FutureNodesTrend} to calculate the expected reduced number of active nodes in a fixed future  period (typically 3 hours) based on the GBDT model prediction. If these two trends are larger than the thresholds ($\xi_H, \xi_P$), \emph{DynamicResourceSleep} is called to reduce the active nodes ($\mathbf{C_A}$) to the current running number ($\mathbf{C_R}$) with a buffer $\sigma$.


In this paper, we exploit the DRS technique on the selected nodes. Alternatively, we can also utilize \emph{CPUfreq governor} and \emph{nvidia-smi} \cite{nvsmi} to adjust the frequency and voltage of CPUs \& NVIDIA GPUs. According to \cite{TangDVFS}, DVFS can not only improve the DL training performance by up to 33\% but also save up to 23\% energy consumption. Evaluations of these techniques will be our future work.


\subsubsection{Evaluation}
We perform similar simulations as \S\ref{sec:qssf-eval} based on the real-world traces from \DCName.
We select a period of 3 weeks (from 1st September to 21st September) in each cluster for evaluation, and the previous records are all used for training the prediction model. Figure \ref{simu_energy} shows the GPU node behaviors in the \texttt{Earth} cluster. Two dashed lines denote the numbers of actual active nodes (blue) and our predictions (orange) respectively. We observe that our prediction can precisely reflect the actual trend with small estimate errors. The red solid line depicts the total number of GPU nodes. It is obvious to see a huge gap between the red and blue lines, representing a large number of energy-wasting idle nodes.
With our CES service, a lot of idle nodes are powered off. The green solid line denotes the number of remaining active nodes. Qualitatively, we observe these nodes are just enough to meet users' demands while significantly reducing the energy waste.


Table \ref{tab_energy} presents a quantitative analysis for each cluster. Our service can remarkably improve the node utilization, especially in \texttt{Earth} (13.0\%) and \texttt{Uranus} (12.6\%), as these clusters are relatively underutilized. Besides, our service only calls the function \emph{NodesWakeUp} 1.1\textasciitilde2.6 times a day in each cluster. In contrast, the vanilla DRS without considering nodes' future trends can incur an average of 34.1 \emph{NodesWakeUp} operations a day, which causes much more turn-on energy overhead and job queue delay. Specifically, assuming each node takes 5 minutes to reboot, our service only affects 251 out of 198k jobs during 21 days. Nevertheless, vanilla DRS leads to nearly 6k jobs affected in the same traces.



Furthermore, we make a rough estimation about the reduction of energy consumption with CES using the data from Table \ref{tab_energy}. The power consumption of one single idle DGX-1 server is around 800 watts (obtained from NVIDIA BMC \cite{bmc} by adding the input values of all PSUs). In addition, the cooling infrastructure typically consumes twice the energy as the servers in datacenter \cite{EnergyModelSurvey}.
Hence, we can save over 1.65 million kilowatt hours of electricity annually across these 4 clusters. This can significantly reduce the operation cost.

\begin{figure}[t]
  \centering
  \includegraphics[width=\linewidth]{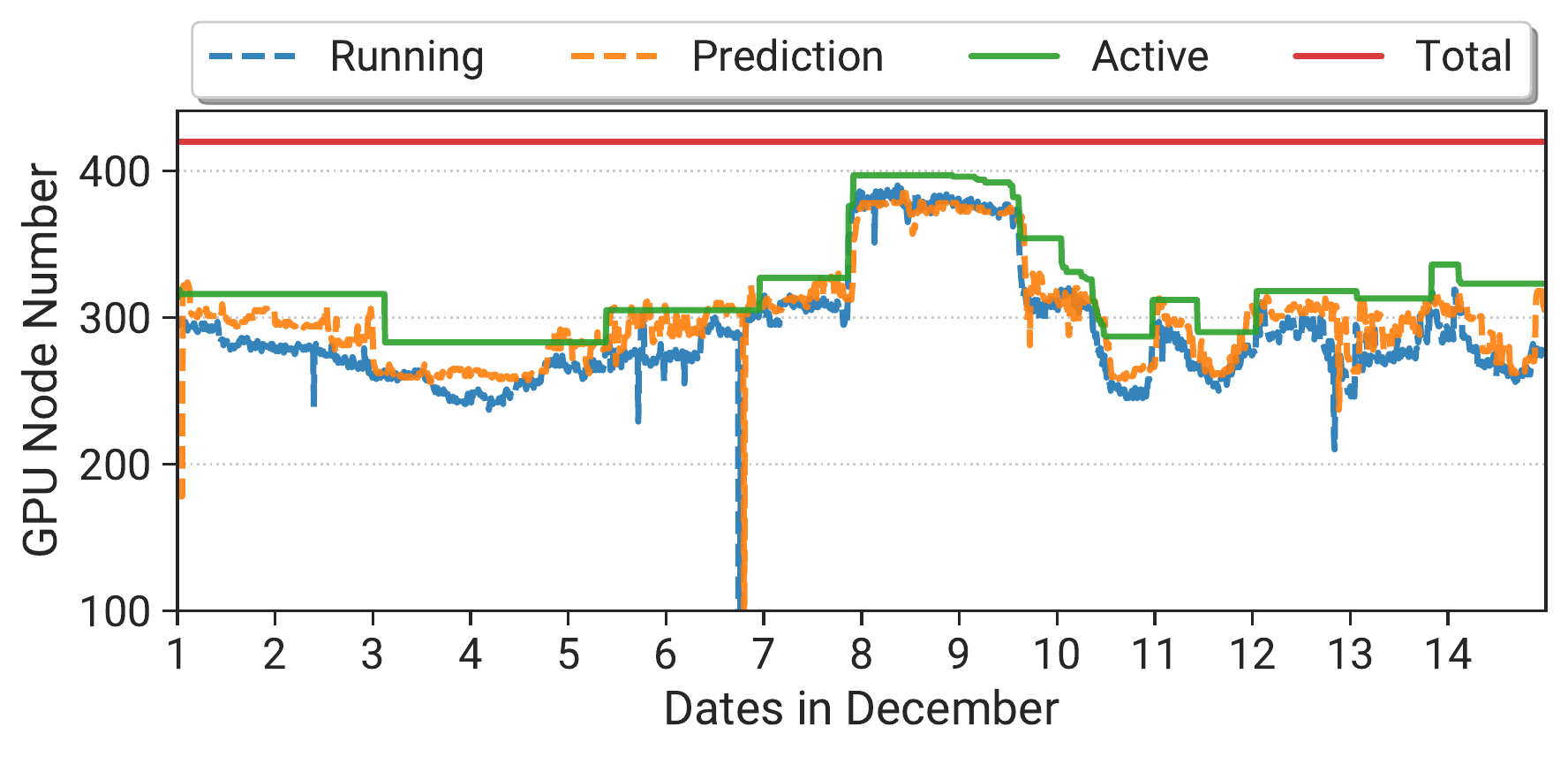}
  \caption{Numbers of compute nodes at different states in Philly from 1st December to 14th December (2 weeks).}
  \Description{philly energy}
  \label{philly_energy}
\end{figure}

We also evaluate our CSE service on the Philly trace. Microsoft provides per-minute statistics about GPU utilization on every node from the Ganglia monitoring system \cite{Philly}, from which we obtain a time sequence about the numbers of total and running GPU nodes. We select a period of 2 weeks (from 1st December to 14th December) for evaluation, and the previous data are all used for training the GBDT model. As shown in Figure \ref{philly_energy}, it is obvious that the change frequency of running nodes in Philly is lower than \texttt{Earth} and its cluster scale is over twice than \texttt{Earth}. Hence, the CES service only needs to take 0.5 times of nodes wake up action on average (Table \ref{tab_energy}), which has a negligible impact on the average JCT. Besides, more than 100 idle nodes can be powered off on average and the cluster node utilization rate is increased from 69\% to 90\% (Table \ref{tab_energy}). This shows the CES service has strong applicability and generality for different clusters and workloads.

\begin{table}[t]
  \caption{Performance of CES service in each cluster of \DCName and Philly.}
  \label{tab_energy}
  \resizebox{\linewidth}{!}{
    \begin{tabular}{@{}ccccc|c@{}}
      \toprule
      \textbf{}                             & \textbf{Venus} & \textbf{Earth} & \textbf{Saturn} & \textbf{Uranus} & \textbf{Philly} \\ \midrule
      Average \# of DRS nodes               & 5.0            & 20.5           & 20.0            & 34.0            & 100.1           \\
      Average times of daily wake up        & 1.1            & 1.3            & 2.0             & 2.6             & 0.5             \\
      Average \# of woken up nodes per time & 6.7            & 6.7            & 7.1             & 9.7             & 27.1            \\ \midrule
      Node utilization (Original)           & 92.7\%         & 82.1\%         & 90.2\%          & 83.8\%          & 69.0\%          \\
      Node utilization (CES)                & 96.2\%         & 95.1\%         & 97.6\%          & 96.4\%          & 90.4\%          \\ \bottomrule
    \end{tabular}}
\end{table}

\section{Related Work}
\textbf{Cluster characterization.}
A number of prior works conducted trace analysis for traditional CPU workloads, e.g., HPC systems \cite{Attig2011, TraceSC20, Rodrigo2018, Joubert2012, Simakov2018, Wolter2006,diversity}, private clouds \cite{Borg, GoogleTrace12, ServerlessATC20, RC}. Amvrosiadis et al. \cite{diversity} presented an analysis of the private and HPC cluster traces about job characteristics, workload heterogeneity, and cluster utilization. They disclosed the correlations among job duration,  behaviors, logical names and user behaviors. We further demonstrate the predictability of cluster resources and workloads information in GPU datacenters.

On the other hand, fewer studies focused on the analysis of characteristics of DL workloads in large-scale clusters. Jeon et al. \cite{Philly} studied the DL job trace from the Microsoft cluster to identify the impact of the DL training job locality on the GPU utilization as well as job failure reasons. Wang et al. \cite{AliPAI} characterized various resource requirements and identified the performance bottlenecks of DL training jobs in Alibaba datacenter.

In comparison, our work provides a more comprehensive analysis about clusters, jobs, and user behaviors. Besides, wec provide more diversity to reflect the latest features of DL algorithms and technologies. These can deepen our understanding about the characteristics of DL clusters and jobs. We uncover seven interesting observations to inspire the new designs of efficient GPU schedulers.

\noindent\textbf{Prediction-based scheduling.}
Prior knowledge of jobs can significantly facilitate the management of cluster resources. Modern cluster systems \cite{BorgK8S, Mesos, YARN} expect users to provide estimates about their job duration, which may not be accurate as the job completion time is unexpected. Several methods were proposed to obtain more precise job information automatically for efficient scheduling. For instance, some schedulers predict the duration of new jobs based on the recurrent 
jobs \cite{ReservationSched, NetworkAware, Hawk,Morpheus}, or job structure knowledge \cite{Dryad,Jockey,ARIA,Apollo, Ernest}. These require the jobs to have explicit periodic patterns or known structures. For more general cases, some systems \cite{Tumanov2016, 3Sigma, RC} predict the estimate from the history of relevant jobs.

For DL training 
job schedulers, Gandiva \cite{Gandiva} leverages intra-job predictability to split GPU-time efficiently across multiple jobs to achieve lower latency. Optimus \cite{Optimus} adopts online fitting to predict model convergence during training. Tiresias \cite{Tiresias} calculates the Gittins index as the job priority according to the duration distribution of previous jobs.  AlloX \cite{AlloX} designs an estimator and profiles jobs in an online manner to determine the job duration. Different from these schedulers only for DL training, our QSSF scheduling service supports all types of DL jobs in the model developing pipeline. This better fits the industry requirements for production GPU clusters.



\noindent\textbf{Energy efficiency for GPU clusters.}
Prior studies \cite{DRS2007, PowerNap} demonstrated that DRS does not affect the performance and efficiency of the cluster, which is consistent with our simulation results. Furthermore, a series of works \cite{TangDVFS, DVFS2011, DVFS2014, GPUPower2016, DVFS2013, MEI2017, DVFS2020} indicated that GPU DVFS has huge benefits to save the energy consumption in GPU clusters. Some job scheduling algorithms \cite{EnergyMei, DVFS2017, mei2021energy} were then proposed based on these techniques to achieve energy efficiency in the clusters. However, they do not consider the impact of cluster future workloads, which can lead to more unnecessary server boot-up delays and extra energy overhead. In our CES service, we further enhance the benefits of the DRS by predicting the cluster usage in advance, which can get better efficiency.


\section{Discussions}

\subsection{Extension to Small-scale Clusters.}
In this paper, we mainly conduct the analysis and characterization of large-scale GPU clusters. Our framework can also be applied to small clusters, for two reasons. (1) In a small cluster, the number of active jobs is also scaled down with the available resources. This does not affect the inherent behaviors and characteristics of jobs and users. (2) We focus on the analysis of VCs (with tens to hundreds of GPUs), which are independent without sharing. They can be regarded as small clusters.

Specifically for our prediction framework, the prediction accuracy does not depend on the cluster size. It is mainly determined by two factors: (1) the size of the training set: more historical data can bring more comprehensive and generalized information, which leads to higher model accuracy; (2) the characteristics of the cluster. QSSF relies on the diverse attributes and distributions of jobs, while CES relies on the seasonal utilization trends of the cluster. So we believe our system can be deployed to small clusters similarly.

\subsection{Future Works}
We identify the following several directions as future work. (1) We aim to design and integrate more services into our framework to make it more comprehensive. (2) Some attributes in our services may not be available in other clusters. We aim to design new qualified models with limited job information for our services. (3) Our services mainly rely on historical job data, and coarse-grained cluster information. We aim to collect and leverage more fine-grained resource information (e.g., GPU memory usage and computation unit utilization, CPU utilization) as features to build more accurate models for better cluster management performance. (4) We are planning to implement our prediction framework in our production clusters, and evaluate its effectiveness at scale.

\renewcommand\UrlFont{\color{blue}\ttfamily\upshape}
\section{Conclusion}

In this paper, we perform a large-scale analysis of the real-world DL job traces from four clusters in our datacenter. We present the characterizations of clusters, jobs and users, and identify seven implications, to guide us to design more efficient GPU cluster systems. Justified by the implication that the behaviors of jobs and clusters are predictable, we introduce a general-purpose GPU cluster management framework, which predicts the future behaviors of jobs and clusters to improve the resource utilization and job performance. As two case studies, we design a QSSF service to improve the average JCT by up to 6.5$\times$, and a CES service to conserve annual power consumption of over 1.65 million kilowatt hours.


\DCName traces are publicly available at \url{https://github.com/S-Lab-System-Group/HeliosData}. We expect they can benefit researchers in the design of GPU datacenter systems.

\renewcommand\UrlFont{\color{black}}
\begin{acks}
  We thank the anonymous reviewers for their valuable comments. This study is supported under the RIE2020 Industry Alignment Fund – Industry Collaboration Projects (IAF-ICP) Funding Initiative, as well as cash and in-kind contributions from the industry partner(s).
\end{acks}


\bibliographystyle{ACM-Reference-Format}
\bibliography{sc21}

\end{document}